\documentclass[a4paper,10pt]{article}

\usepackage{multirow}	
\usepackage[latin1]{inputenc}
\usepackage[T1]{fontenc}
\usepackage[british]{babel}
\usepackage{graphicx}
\usepackage[dvips]{geometry}
\usepackage{float}
\usepackage{array}
\usepackage{amsmath}
\usepackage{amsfonts}
\usepackage{tikz}
\usepackage{psfrag}
\usepackage{graphicx}
\usepackage{amssymb}
\usepackage{amsmath}
\usepackage{leftidx} 

\newcommand{\ket}[1]{\left|#1  \right>}
\newcommand{\bra}[1]{\left<#1  \right|}

\def\be{\begin{equation}}
\def\ee{\end{equation}}
\def\ba{\begin{eqnarray}}
\def\ea{\end{eqnarray}}

\newcommand{\ua}{\uparrow}
\newcommand{\da}{\downarrow}

\newcommand{\ve}[2]{
\begin{scope}[xshift=#1 cm,yshift=#2 cm,rotate=45,scale=0.500]
	\clip[draw] (-0.5,-0.5) rectangle (0.5,0.5);
	\draw[line width=3pt] (-0.5,-0.5) circle (0.5cm)  (0.5,0.5) circle (0.5cm);
	\draw (-0.5,-0.5) rectangle (0.5,0.5);
\end{scope}
}

\newcommand{\vi}[2]{
\begin{scope}[xshift=#1 cm,yshift=#2 cm,rotate=45,scale=0.500]
	\clip[draw] (-0.5,-0.5) rectangle (0.5,0.5);
	\draw[line width=3pt] (-0.5,0.5) circle (0.5cm)  (0.5,-0.5) circle (0.5cm);
	\draw (-0.5,-0.5) rectangle (0.5,0.5);
\end{scope}
}

\newcommand{\ves}[2]{
\begin{scope}[xshift=#1 cm,yshift=#2 cm,rotate=45,scale=0.707]
	\clip[draw] (-0.5,-0.5) rectangle (0.5,0.5);
	\draw[line width=3pt] (-0.5,-0.5) circle (0.5cm)  (0.5,0.5) circle (0.5cm);
	\draw (-0.5,-0.5) rectangle (0.5,0.5);
\end{scope}
}

\newcommand{\vis}[2]{
\begin{scope}[xshift=#1 cm,yshift=#2 cm,rotate=45,scale=0.707]
	\clip[draw] (-0.5,-0.5) rectangle (0.5,0.5);
	\draw[line width=3pt] (-0.5,0.5) circle (0.5cm)  (0.5,-0.5) circle (0.5cm);
	\draw (-0.5,-0.5) rectangle (0.5,0.5);
\end{scope}
}

\newcommand{\vesg}[2]{
\begin{scope}[xshift=#1 cm,yshift=#2 cm,rotate=45,scale=0.707]
	\filldraw[gray] (-0.5,-0.5) rectangle (0.5,0.5);
	\clip[draw] (-0.5,-0.5) rectangle (0.5,0.5);
	\draw[line width=3pt] (-0.5,-0.5) circle (0.5cm)  (0.5,0.5) circle (0.5cm);
	\draw (-0.5,-0.5) rectangle (0.5,0.5);
\end{scope}
}

\begin{document}

\title{Conformal field theory at central charge $c=0$: \\
a measure of the indecomposability ($b$) parameters.}

\author{J\'er\^ome Dubail$^{1,2}$, Jesper Lykke Jacobsen$^{2,1}$ and Hubert Saleur$^{1,3}$ \\
[2.0mm]
  ${}^1$Institut de Physique Th\'eorique, CEA Saclay,
  91191 Gif Sur Yvette, France \\
  ${}^2$LPTENS, 24 rue Lhomond, 75231 Paris, France \\
  ${}^3$Department of Physics,
  University of Southern California, Los Angeles, CA 90089-0484}


\maketitle

\begin{abstract}
	A good understanding of conformal field theory  (CFT) at $c=0$ is vital to the physics of disordered systems, as well as geometrical problems such as polymers and percolation. Steady progress has shown that these CFTs should be logarithmic, with indecomposable operator product expansions, and indecomposable
	representations of the Virasoro algebra.  In one of the earliest papers on the subject, V. Gurarie introduced a single parameter $b$ to quantify this indecomposability in terms of the logarithmic partner $t$ of the stress energy tensor $T$. He and A. Ludwig conjectured  further that $b=-\frac{5}{8}$ for polymers and $b=\frac{5}{6}$ for percolation. While a lot of physics may be hidden behind this parameter - which has also given rise to a lot of discussions - it had remained very elusive up to now, due to the lack of available methods to measure it experimentally or numerically, in contrast say with the central charge. We show in this paper how to overcome the many difficulties in trying to measure $b$. This requires control of a lattice scalar product, lattice Jordan cells, together with a precise construction of the state $L_{-2}|0\rangle$. The final result is that $b=\frac{5}{6}$ for polymers. For percolation, we find that $b=-\frac{5}{8}$ within an XXZ or supersymmetric representation. In the  geometrical representation, we do not find a Jordan cell for $L_0$ at level two (finite-size Hamiltonian and transfer matrices are fully diagonalizable), so there is no $b$ in this case. 
\end{abstract}



\section*{Introduction}

In the last twenty years or so since the seminal paper \cite{BPZ} of
Belavin, Polyakov and Zamolodchikov, Conformal Field Theory (CFT) has
proven amazingly successful. It is now an essential item in the
toolbox of condensed matter and string theorists, and has had a
profound impact on several sub-fields of modern mathematics.

Yet, despite this flurry of successes, some very fundamental questions
have remained unanswered to this day. One of these questions concerns
percolation, the very geometrical critical problem where CFT has
obtained some of its most impressive results. In a nutshell, despite
years of work and some progress (see below), we do not know any
conformal field theory describing at least some of the geometrical
observables (be it hulls, clusters or backbones) in a fully consistent
way. As a result, many quantities involving bulk correlations
functions---for instance the equivalent of the Binder cumulant
\cite{Binder} in this problem \cite{SaleurDerrida}---are, to this day,
unknown analytically.
 
Another vexing question concerns the celebrated transition between
plateaux in the integer quantum Hall effect. The evidence is strong
that it corresponds to quantum critical points of the 2+1 dimensional
electron gas. The physics of the transition is somewhat well
understood (and involves an interplay between disorder, which tends to
localize the electrons, and the kinetic energy, quenched by the strong
magnetic field, which causes delocalization), and qualitatively well
described by a 2D sigma model with topological term \cite{Pruisken}.
Nevertheless, a precise identification of the low energy effective
field theory is still lacking to this day, despite the wealth of
theoretical and numerical works on the topic, and the fact that this
theory is expected to be conformally invariant (see \cite{Zirnbauer}
for an insightful review).

The origin of these difficulties is, ultimately, the fact that the
CFTs describing these problems have to be non unitary, with vanishing
central charge. Non-unitarity in a CFT can have rather mild
consequences---like in the minimal theories such as the Yang Lee edge
singularity. It took a while to realize that non-unitarity would most
often (in particular, when $c=0$) imply in fact indecomposability,
leading to a very difficult kind of theory called a logarithmic CFT
(LCFT).

LCFTs were probably first encountered in published form in a paper by
Rozansky and Saleur \cite{RozanskySaleur}. In this paper, the authors
studied a particular kind of $c=0$ theory with $U(1|1)$ supergroup
symmetry, and stumbled upon four-point functions involving logarithmic
dependence on the cross ratio. These authors correctly related this
property to the indecomposability of the operator product expansions (OPE),
and to the non-diagonalizability of the $L_0$ generator, inherited
from the non semi-simplicity of the symmetry algebra. Shortly after,
Gurarie \cite{Gurarie1} pointed out that these features were in fact
necessary to have a consistent, non-trivial CFT at $c=0$. Later,
Gurarie \cite{Gurarie2} and Gurarie and Ludwig \cite{GurarieLudwig}
built up a very attractive formalism within which CFTs at $c=0$ must
possess, in addition to their stress energy tensor $T(z)$, an extra
field whose holomorphic part, $t(z)$, has conformal weight two. The
singular part of the OPE between $T(z)$ and $t(z)$ is determined up to
a new universal number, an ``anomaly'' usually denoted by $b$. This
parameter is expected in \cite{Gurarie2,GurarieLudwig} to play a very
important physical role. It might obey a ``c-theorem'' \cite{Gurarie1}
and thus indicate possible directions of RG flows within theories with
vanishing central charge. Its value also dictates the existence of
null vectors for conformal weights contained in the Kac table, and is
thus profoundly related with the determination of four-point
functions.  Gurarie and Ludwig \cite{GurarieLudwig} suggested, based
on knowledge of critical exponents from Coulomb gas calculations and
some heuristic hypotheses, that $b=-\frac{5}{8}$ for polymers and
$b=\frac{5}{6}$ for percolation.%
\footnote{We shall define $b$ precisely below.}

Despite the appeal of the ideas proposed in
\cite{Gurarie2,GurarieLudwig}, only very little progress has happened
to make them into a powerful theoretical tool. It is not clear for
instance to what extent it is possible or helpful to study further the
extension of the Virasoro algebra proposed in
\cite{GurarieLudwig}. Considerable difficulties also arise when one
tries to build non-chiral theories by combining the left and right
moving sectors. Most examples worked out so far indeed concern only
boundary LCFTs.

Meanwhile, the subject has matured with the understanding that,
probably, rather than a single parameter $b$, LCFTs might be
characterized by a complicated structure of indecomposable Virasoro
modules. This structure involves maps between sub-modules, and several
parameters to describe their precise action. This line of thought
shows how close the problem is to the theory of non semi-simple Lie
(super)algebras, whose representation theory is more often than not of
the ``wild'' type. Progress has been steady in trying to understand
indecomposable Virasoro modules \cite{Rohsiepe,
  KauschGaberdiel,MathieuRidout, MathieuRidout1, KytolaRidout} in
general LCFTs. For $c=0$ the $b$ parameters appear indeed as some
particular coefficient describing the embedding of a sub-module. Some
simple algebraic arguments suggest then that $b=-\frac{5}{8}$ for
percolation and $b=\frac{5}{6}$ for polymers, that is, the value of
Gurarie and Ludwig \cite{GurarieLudwig} {\sl up to a switch}.

In the last two years, more progress has occurred from the direction
of lattice models. It had long been known that the representation
theory of lattice algebras (mostly, the Temperley-Lieb algebra) bore
some striking resemblance to that of the Virasoro algebra. In
particular, it had been known that when $q$ is a root of unity, the
hamiltonian (the lattice discretization of $L_0$, or $L_0+\bar{L}_0$
depending on the boundary conditions) can sometimes be
non-diagonalizable, and that Jordan cells appear, very much mimicking
the ones expected in the continuum limit. This was explored in a much
more systematic fashion in works by Pearce, Rasmussen and Zuber
\cite{PRZ,RP,RP1} and independently by Read and Saleur
\cite{ReadSaleur}. The outcome of these analyses was a rather coherent
picture of the indecomposable Virasoro modules appearing in some
families of boundary LCFTs, including percolation and polymers,
together with some formal but potentially useful results about
fusion. This picture was interpreted in interesting mathematical terms
in \cite{Azat}.

Yet, large parts of these constructions are still speculative. For
instance, the lattice analysis clearly shows which modules are mapped
into each other under the action of the lattice algebras, but what
this becomes precisely in the continuum limit---including which
Virasoro generators and states are involved in these mappings,
together with precise values of the corresponding coefficients---is
largely conjectural. Moreover, it is clear that only the `simplest
indecomposable' modules have been encountered so far, and one is still
far from being able to guess what would happen, say, for the quantum
Hall plateau transition.

One of the crucial remaining obstacles is that the features that make
LCFTs difficult are also very hard to observe or measure directly.  It
is quite striking for instance, that despite the 15 year old
controversy around the value of the $b$ parameter for percolation or
polymers---problems which are usually rather easy to study
numerically---no way to measure this parameter has been available, up
to now. One of the only forays in this direction is the paper by Koo
and Saleur \cite{KooSaleur}, who attempted to define and study a
straightforward regularization of the Virasoro algebra on the
lattice. Their success was only partial, and hampered by the fact that
the continuum limit of commutators is not the commutator of the
continuum limits.

We shall report in this paper a method to measure the parameter $b$ in
lattice models. While it might not be general enough to work for all
theories at $c=0$, it certainly will allow us to assert what the
values of this parameter for percolation and polymers actually are.

The paper is organized as follows. In the first section, we recall
basic facts about the $b$ number in LCFTs at $c=0$, and define a
general strategy to measure it. This strategy requires overcoming two
difficulties. One is finding a scalar product on the lattice that goes
to the Virasoro-Shapovalov form in the continuum limit.  The other is
finding a regularization of the $L_{-2}$ operator---or more precisely,
of its action on the vacuum.  These difficulties are solved in the
next two sections. In section 4, we finally gather all the pieces to
study the Jordan cells for percolation and polymers, and measure $b$.

\section{The $b$ number}

In a nutshell, the argument of Gurarie
\cite{Gurarie1,Gurarie2,GurarieLudwig} went as follows. Conformal
invariance fixes the coefficient of the stress tensor $T$ in the OPE
of an operator with itself (assuming there is a single field with
$h=2,\bar{h}=0$) to be of the form
\begin{equation}
\label{phiphiOPE}
\phi(z)\phi(0)\approx a_\phi z^{-2h_\phi}\left(1+\frac{2h_\phi}{c}T(z)+\ldots\right)
\end{equation}
Here $a_\phi$ is an amplitude, determining the normalization of the two-point function 
\begin{equation}
\langle \phi(z,\bar{z})\phi(0,0)\rangle=|a_\phi|^2 \frac{1}{z^{2h_\phi}\bar{z}^{2\bar{h}_\phi}}
\end{equation}
If one want to keep $a_\phi$ finite in a theory at $c=0$ for an operator with $h_\phi\neq 0$, the $\frac{2 h_\phi}{c}$ factor poses problems as it diverges. 

There are various ways to resolve the difficulty (this was also
discussed in \cite{KoganNichols,Cardylog}. First, a divergence in the
OPE coefficient (which would manifest itself in a physical quantity
such as a four point function) might not be a problem after all, but
have a physical meaning, related with some $n\to 0$ limit. Second, it
could be that demanding $a_\phi$ finite is not physically meaningful,
again in relation with an $n\to 0$ limit. Recall that since we are
dealing with non-unitary theories, having a vanishing two-point
function does not mean that the corresponding operator is zero. The
third way to resolve the difficulty is to admit that there are other
operators with $h=2$ appearing on the right-hand side of the OPE
(\ref{phiphiOPE}). These operators might be part of a supersymmetry
multiplet, such as those that might occur in supergroup WZW models
\cite{SchomerusSaleur} and other supersymmetric CFTs
\cite{ReadSaleur01}. We note that in such cases, there is no need for
a logarithmic dependence in the OPE at this order, although the theory
will in general be logarithmic (see, e.g., \cite{SchomerusSaleur}.  The
other scenario proposed in \cite{GurarieLudwig} is of an $n\to 0$
limit, with another operator whose dimension is generically different
from 2 becoming degenerate with $T$. In this case---examples of which
can be worked out in details for the models with central charge
$c=1-\frac{6}{x(x+1)}$, in the limit $x\to 2$---one expects the OPE of $\phi$
with itself to read instead
\begin{equation}
\phi(z)\phi(0)\approx a_\phi z^{-2h_\phi}\left(1+\frac{2h_\phi}{b}z^2[t(0)+\log(z) T(0)]+\ldots\right) \,.
\end{equation}
The operator $t$ is called `logarithmic partner' of $T$, and forms with it a logarithmic pair, that is, a set of fields on which the $L_0$ operator is not diagonalizable but has the form of a Jordan cell
\begin{equation}
L_0| T\rangle=2|T\rangle,~~~L_0 |t\rangle=2|t\rangle+|T\rangle\label{Tcell}
\end{equation}
In terms of two-point functions one has
\begin{subequations}
\begin{eqnarray}
\langle T(z)T(0)\rangle&=&0 \\
\langle T(z)t(0)\rangle&=&\frac{b}{z^4} \\
\langle t(z)t(0)\rangle&=&\frac{-2b\log z+a}{z^4}
\end{eqnarray}
\label{corrf}
\end{subequations}
where the last result comes from imposing global conformal invariance \cite{GurarieLudwig}. Note that the {\sl normalization} of $T$ is crucial in defining the numerical value of $b$, once a choice has been made that the $L_0$ Jordan cell has off-diagonal term unity (which translates into the relative normalization of the last two equations).  There remains, on the other hand, a choice $t\to t+\hbox{constant } T$, which does not affect  the leading terms in the OPE, but only the constant $a$. 

The universality of these OPEs, the value of the number $b$ and its properties (e.g., under RG flows) have all been subject of intense debate. It is not our purpose to review this debate in detail. We only wish to recall that, when thinking of the limit of minimal models as $c\to 0$, one is naturally led to two candidates which might become degenerate with $T$: the $(3,1)$ field and the $(1,5)$ field (in Kac table notations). The former choice leads to $b=\frac{5}{6}$, and the latter to $b=-\frac{5}{8}$. Using a variety of arguments, Gurarie and Ludwig \cite{GurarieLudwig} predicted that the former choice corresponds to percolation, and the latter to polymers, which are therefore profoundly different logarithmic CFTs. 

While one can dwell at great length on the merits of the arguments in \cite{GurarieLudwig} we believe it is more constructive to ask whether one can, in fact, measure the parameter $b$. Some considerable difficulties arise when thinking of this possibility. One might, of course, try to identify $t$ in a lattice model of percolation or polymers. After all, it is quite possible to get a handle on the meaning of $T$ itself \cite{KadanoffCeva,Benjamin}. But measuring directly correlation functions such as those in (\ref{corrf}) is a very daunting task, not to mention the difficulty of extracting properly a logarithmic dependence. We also note that 
the formalism in \cite{GurarieLudwig} is really a chiral one, where the job of glueing back together the left and right parts of the theory is left for further work, and might lead to considerable surprises. This chiral aspect naturally suggests turning to the physics of the boundary theory, which is in fact what is mostly considered in \cite{GurarieLudwig}. 

Short of measuring correlation functions on the lattice, the next best hope would be to access numerically $b$ through a lattice analog of the Jordan cell structure in ($\ref{Tcell}$). Some progress in reproducing algebraic relations---e.g., the Virasoro commutation relations---in lattice models was reported in fact in \cite{KooSaleur}, and it is a natural idea to try to push further the results obtained there. Introduce first the scalar product and the notion of adjoint from in and out states as usual   \cite{Flohr}, with
 $L_n^\dagger=L_{-n}$ (of course now the form is not positive definite). Then the relations equivalent to (\ref{corrf}) are
\begin{equation}
L_0=\left(\begin{array}{cc}
2&1\\
0&2\end{array}\right)
\end{equation}
in the basis $\{T,t\}$, where
\begin{eqnarray}
|T\rangle=L_{-2}|0\rangle&,& \langle T|T\rangle=0\nonumber\\
L_2|t\rangle= b |0\rangle&,& \langle t|T\rangle=b \,.
\end{eqnarray}
We see that a possible strategy to measure $b$ would involve finding a Jordan cell for the hamiltonian (the lattice version of $L_0$) in finite size. One would then have to identify the states corresponding to $T,t$ and normalize them properly. Finally one would have to calculate the scalar product $\langle t|T\rangle$.  We shall accomplish this program here, but first let us stress why it is difficult, and has not been done before.

The first obstacle is to find a Jordan cell in lattice models. This in fact is quite easy, as has been known for a while. The point is that the `hamiltonians' in lattice models whose continuum limit is a CFT are often non-diagonalizable, a consequence of their deep relationship with lattice symmetries (quantum groups) or lattice algebras (Temperley-Lieb algebras) which have non semi-simplicity properties fully emulating those of the continuum limit. Nevertheless, the question whether there exists a Jordan cell for $L_0$ in a given lattice model is not easily answered a priori---we shall mention some surprises along these lines later. 

Having obtained the Jordan cell, one needs to identify $T$. While it is natural to think that $T$ is an eigenvector of the hamiltonian, a very important obstacle is its normalization. While crucial to determine the value of $b$, it cannot be done via a standard eigenvector normalization since the two-point function of $T$ vanishes in the continuum, which manifests itself by the fact that $|T\rangle$---or rather its lattice equivalent---is a null state. To normalize $T$, we shall resort to a very indirect trick that we call the ``trousers trick'', which is really at the heart of our solution of the problem.

Finally, one needs to properly normalize $t$ via the desired Jordan cell structure. This is  quite involved with big transfer matrices, and requires some numerical tricks. We note in passing that, according to ($\ref{corrf}\, c$), $|t\rangle$ turns out to have infinite norm square.

\section{Lattice scalar products}

\subsection{Hamiltonians and Transfer Matrices}
A 2D classical statistical model can be viewed as a 1+1D model
evolving in imaginary time. At large length scales, the low-energy
excitations of the latter are described by some quantum field
theory. When the 2D model is at a critical point, these excitations
are gapless, hence the field theory becomes conformal. A 1+1D system
of width $L$ is defined by its Hamiltonian $H_L$, or
by its transfer matrix $T_L$ when it comes from a 2D statistical model.

Let us start by considering the Ising model as an example. The critical Ising chain
is defined by the Hamiltonian (say with periodic boundary conditions
$\sigma^z_{L+1} = \sigma^z_1$)
\begin{equation}
\label{eq:HamIsing}
H_L = -\sum_{i=1}^{L} \sigma_i^z \sigma_{i+1}^z - \sum_{i=1}^{L} \sigma_i^x
\end{equation}
acting on the space of spin configurations $\ket{\uparrow \downarrow
\downarrow \uparrow \dots \downarrow}$. $H_L$ is hermitian, therefore
in the continuum limit is described by a unitary CFT. It is well-known
that it corresponds to the minimal model $\mathcal{M}_{4,3}$ with central
charge $c=1/2$.

Now, our second example will turn out to be non-unitary.
It is a dense loop model, given by the
transfer matrix (in the picture $L=6$)
\begin{equation}
	\begin{tikzpicture}[scale=0.75]
		\draw (-2,0.5) node{$T_L \; =$};
		\draw[rotate=45] (0,0) rectangle ++(0.5,0.5);
		\draw[rotate=45] (0.5,0) rectangle ++(0.5,0.5);
		\draw[rotate=45] (0.5,-0.5) rectangle ++(0.5,0.5);
		\draw[rotate=45] (1,-0.5) rectangle ++(0.5,0.5);
		\draw[rotate=45] (1,-1) rectangle ++(0.5,0.5);
		\draw[rotate=45] (1.5,-0.5) -- ++(0.5,0);
		\draw[rotate=45] (0.5,0.5) -- ++(0,0.5);
		\draw[dashed,scale=1.414] (-0.25,0) -- ++(0,0.85);
		\draw[dashed,scale=1.414] (1.25,0) -- ++(0,0.85);
        \end{tikzpicture}
\end{equation}
where each square is a sum over two terms
$$
	\begin{tikzpicture}[scale=0.75]
			\draw[xshift=-8, yshift=-10, rotate=45] (0,0) rectangle ++(0.5,0.5);
			\vi{1}{0}
			\ve{2.2}{0}
			\draw (1.62,0) node{$+$};
			\draw (0.3,0) node{$\equiv$};
	\end{tikzpicture}
$$
and one takes periodic boundary conditions (dotted lines
are identified). Iterations of $T_L$ (from bottom to top)
build the partition function of the dense loop model on a cylinder.

We have not specified yet on which configuration space $T_L$
is acting. Here the basis states will encode the planar pairings of
$L$ points, such as
$\ket{\; \begin{tikzpicture}
\draw[thick] (0,0) arc (-180:0:0.1 and 0.25);
\draw[thick] (0.4,0) arc (-180:0:0.3 and 0.25);
\draw[thick] (0.6,0) arc (-180:0:0.1 and 0.15);
\end{tikzpicture}\;}$ or $\ket{\; \begin{tikzpicture}
\draw[thick] (0,0) arc (-180:0:0.5 and 0.25);
\draw[thick] (0.2,0) arc (-180:0:0.1 and 0.15);
\draw[thick] (0.6,0) arc (-180:0:0.1 and 0.15);
\end{tikzpicture}\;}$
for $L=6$. A state in this model is a linear combination of such basis
states. The action of $T_L$ on basis states is found by stacking the
diagrams from bottom to top, e.g.:
$$
\begin{tikzpicture}
	\begin{scope}
		\clip (-0.353,-0.4) rectangle (1.767,0.8);
		\vi{0}{0}
		\ve{0.707}{0}
		\vi{1.414}{0}
		\ve{0.353}{0.353}
		\vi{1.06}{0.353}
		\vi{-0.353}{0.353}
		\vi{1.767}{0.353}
	\end{scope}
	\draw[dashed] (-0.353,-0.3) -- (-0.353,0.8); 
	\draw[dashed] (1.767,-0.3) -- (1.767,0.8);
	\draw (2.75,0.2) node{$\ket{\; \begin{tikzpicture}
	\draw[thick] (0,0) arc (-180:0:0.1 and 0.25);
	\draw[thick] (0.4,0) arc (-180:0:0.3 and 0.25);
	\draw[thick] (0.6,0) arc (-180:0:0.1 and 0.15);
	\end{tikzpicture}\;} = $};
	
	\begin{scope}[xshift=4.2cm,yshift=0.2cm]
		\begin{scope}
		\clip (-0.353,-0.4) rectangle (1.767,0.8);
		\vi{0}{0}
		\ve{0.707}{0}
		\vi{1.414}{0}
		\ve{0.353}{0.353}
		\vi{1.06}{0.353}
		\vi{-0.353}{0.353}
		\vi{1.767}{0.353}
		\end{scope}
		\draw[dashed] (-0.353,-0.3) -- (-0.353,0.8); 
		\draw[dashed] (1.767,-0.3) -- (1.767,0.8);
		\draw[thick] (0.5,-0.4) arc (-180:0:0.55 and 0.35);
		\draw[thick] (0.85,-0.4) arc (-180:0:0.2 and 0.25);
		\draw[thick] (-0.2,-0.4) arc (-180:0:0.2 and 0.3);
	\end{scope}
	
	\draw (7,0.2) node{$= \ket{\; \begin{tikzpicture}
	\draw[thick] (0,0) arc (-180:0:0.3 and 0.25);
	\draw[thick] (0.2,0) arc (-180:0:0.1 and 0.15);
	\draw[thick] (0.8,0) arc (-180:0:0.1 and 0.25);
	\end{tikzpicture}\;}$};
\end{tikzpicture}
$$
Each closed loop gets a Boltzmann weight $n$.
For our formalism to be complete, we need to define a transposition
operation (or equivalently a scalar product).

\subsection{Transposition and scalar product}

How can we define the forms $\left< . \right|$\;? In our geometrical
formulation, the elements of the transfer matrix should act on
such an object from top to bottom:
$$
\begin{tikzpicture}
	\begin{scope}
		\clip (-0.353,-0.4) rectangle (1.767,0.8);
		\vi{0}{0}
		\ve{0.707}{0}
		\vi{1.414}{0}
		\ve{0.353}{0.353}
		\vi{1.06}{0.353}
		\vi{-0.353}{0.353}
		\vi{1.767}{0.353}
	\end{scope}
	\draw[dashed] (-0.353,-0.3) -- (-0.353,0.8); 
	\draw[dashed] (1.767,-0.3) -- (1.767,0.8);
	\draw (-1.25,0.2) node{$\left< {\; \begin{tikzpicture}
\draw[thick] (0,0) arc (180:0:0.3 and 0.25);
\draw[thick] (0.2,0) arc (180:0:0.1 and 0.15);
\draw[thick] (0.8,0) arc (180:0:0.1 and 0.25);
\end{tikzpicture}}\;\right|$};
	
	\begin{scope}[xshift=2.95cm,yshift=-0.05cm]
		\draw (-0.75,0.2) node{$=$};
		\begin{scope}
		\clip (-0.353,-0.4) rectangle (1.767,0.8);
		\vi{0}{0}
		\ve{0.707}{0}
		\vi{1.414}{0}
		\ve{0.353}{0.353}
		\vi{1.06}{0.353}
		\vi{-0.353}{0.353}
		\vi{1.767}{0.353}
		\end{scope}
		\draw[dashed] (-0.353,-0.3) -- (-0.353,0.8); 
		\draw[dashed] (1.767,-0.3) -- (1.767,0.8);
		\draw[thick] (1.2,0.75) arc (180:0:0.2 and 0.3);
		\draw[thick] (0.15,0.75) arc (180:0:0.2 and 0.25);
		\draw[thick] (-0.2,0.75) arc (180:0:0.55 and 0.35);
	\end{scope}
	
	\draw (6.2,0.15) node{$= \; n \; \left< {\; \begin{tikzpicture}
\draw[thick] (0,0) arc (180:0:0.1 and 0.25);
\draw[thick] (0.4,0) arc (180:0:0.1 and 0.25);
\draw[thick] (0.8,0) arc (180:0:0.1 and 0.25);
\end{tikzpicture}}\;\right|$};
\end{tikzpicture}
$$
The \textit{loop scalar product} is then defined
for basis states by glueing the mirror image of the first state
on top of the second one, giving a weight $n$ per closed loop.
For instance
\begin{equation}
\label{loopscalar}
\begin{tikzpicture}
\begin{scope}[xshift=-3.4cm]
	\draw (0,0) node{$\left< {\; \begin{tikzpicture}
	\draw[thick] (0,0) arc (-180:0:0.1 and 0.25);
	\draw[thick] (0.4,0) arc (-180:0:0.3 and 0.25);
	\draw[thick] (0.6,0) arc (-180:0:0.1 and 0.15);
	\end{tikzpicture}}
	\; | \; {\begin{tikzpicture}
	\draw[thick] (0,0) arc (-180:0:0.5 and 0.25);
	\draw[thick] (0.2,0) arc (-180:0:0.1 and 0.15);
	\draw[thick] (0.6,0) arc (-180:0:0.1 and 0.15);
	\end{tikzpicture}} \; \right>$};
\end{scope}
\draw (-1.7,-0.03) node{$\equiv$};
\begin{scope}
	\draw (0,0) node{$\left< {\; \begin{tikzpicture}
	\draw[thick] (0,0) arc (180:0:0.1 and 0.25);
	\draw[thick] (0.4,0) arc (180:0:0.3 and 0.25);
	\draw[thick] (0.6,0) arc (180:0:0.1 and 0.15);
	\end{tikzpicture}}
	\; | \; {\begin{tikzpicture}
	\draw[thick] (0,0) arc (-180:0:0.5 and 0.25);
	\draw[thick] (0.2,0) arc (-180:0:0.1 and 0.15);
	\draw[thick] (0.6,0) arc (-180:0:0.1 and 0.15);
	\end{tikzpicture}} \; \right>$};
\end{scope}
\draw (1.75,-0.03) node{$=$};
\begin{scope}[xshift=2.15cm]
	\draw[thick] (0,0) arc (180:0:0.1 and 0.25);
	\draw[thick] (0.4,0) arc (180:0:0.3 and 0.25);
	\draw[thick] (0.6,0) arc (180:0:0.1 and 0.15);
	\draw[thick] (0,0) arc (-180:0:0.5 and 0.25);
	\draw[thick] (0.2,0) arc (-180:0:0.1 and 0.15);
	\draw[thick] (0.6,0) arc (-180:0:0.1 and 0.15);
\end{scope}
\draw (3.75,0.05) node {$= \; n^2$};
\end{tikzpicture}
\end{equation}
This definition is extended by linearity to all the other
states. Note that
$\ket{l} \equiv \ket{\; \begin{tikzpicture}
\draw[thick] (0,0) arc (-180:0:0.1 and 0.25);
\draw[thick] (0.4,0) arc (-180:0:0.3 and 0.25);
\draw[thick] (0.6,0) arc (-180:0:0.1 and 0.15);
\end{tikzpicture}\;} - \ket{\; \begin{tikzpicture}
\draw[thick] (0,0) arc (-180:0:0.5 and 0.25);
\draw[thick] (0.2,0) arc (-180:0:0.1 and 0.15);
\draw[thick] (0.6,0) arc (-180:0:0.1 and 0.15);
\end{tikzpicture}\;}$
has norm squared $\left< l | l \right> = 2 n^2 (n-1)$, which is negative if
$n<1$. So for generic $n$ the loop scalar product is not positive definite. There is no reason to expect otherwise, as the theories we are dealing with are not unitary.
We note that this scalar product is well-known in the theory of the Temperley-Lieb algebra \cite{Jones}.

\paragraph{}
For the Ising chain ($\ref{eq:HamIsing}$), the scalar product is 
much simpler. It is the usual (Euclidian) one $\leftidx{_L}{\left< s_1 \dots s_L
| s'_1 \dots s'_L \right>}{_L} = \delta_{s_1,s'_1} \dots \delta_{s_L, s'_L}$
where $s_i,s'_i \in \left\{ \uparrow ,\downarrow \right\}$. In general,
if there is a unitary
representation of the model, one can find a basis in which the scalar
product above is just the Euclidean one. For example,
 in minimal models the RSOS height
representation provides an orthonormal
basis $\ket{h}_L = \ket{ \{h_i\} }_L$, so writing
the states as $\ket{a}_L = \sum_{h} a_h \ket{h}_L$ the
scalar product is simply $\leftidx{_L}{\left<a | b \right>}{_L} = \sum_{h}
a_{h} b_{h}$. In non-unitary models, the simple loop model above shows that
states and scalar product can be more subtle.

\paragraph{}

At a critical point the Hamiltonian can be related to the CFT
translation operator along the cylinder $L_0 + \bar{L}_0 -c/12$
\cite{Affleckc,Cardyc}
\begin{equation}
 \label{finitesize_cylinder}
  H_L = E_{0} L + \frac{2 \pi v_F}{L} \left(L_0 + \bar{L}_0 -\frac{c}{12}\right)
  + \mathcal{O} \left( L^{-2} \right) \,,
\end{equation}
where $v_F$ denotes the Fermi velocity in the dispersion relation $E
\sim v_F \left| k \right|$, and $E_0 L$ is the (non-universal)
extensive part of the ground state energy. A similar relation holds
for $\log T_L$, since $T_L \simeq {\rm e}^{-a H_L}$, where $a$
is the lattice spacing.

In finite size, when the Hamiltonian (or equivalently the transfer matrix)
is diagonalizable, one can
label the right eigenstates as $\ket{0}_L, \; \ket{1}_L, \; \ket{2}_L, \dots$
with $E_{0,L} \leq E_{1,L} \leq E_{2,L} \leq \dots$. The knowledge
of these energies for different sizes $L$ is a very useful way
of extracting the conformal spectrum using ($\ref{finitesize_cylinder}$)
\cite{Affleckc,Cardyc}. We claim in this paper that we can go further than
the spectrum, and that scalar products can be measured as well.
Let us introduce the left eigenstates of $H_L$:
$\leftidx{_L}{\bra{0}}{}, \; \leftidx{_L}{\bra{1}}{},\;
\leftidx{_L}{\bra{2}}{}, \dots$. We expect these different states to correspond
to the conformal states $\ket{0}, \; L_{-2} \ket{0} , \; \bar{L}_{-2} \ket{0},
\; L_{-2} \bar{L}_{-2} \ket{0}, \dots$ (of course, descendants of other
primary operators are also present in the spectrum, depending
on the model we are looking at). 
We note that for different energies $E_i \neq E_j$ (or even different
momenta) one has $\leftidx{_L}{\left< i | j \right>}{_L} =0$, like in the continuum limit
 where scalar products between states at different levels
are zero: $\left< 0 | L_{k'_1} \dots L_{k'_m} L_{-k_n}
\dots L_{-k_1} | 0 \right> =0 \;$ if $\;\sum_{i=1}^m k'_i \neq \sum_{i=1}^n k_i$.

In general, one can mimic on the lattice the construction of in and out states and scalar products of the continuum 
limit \cite{Ginsparg,Flohr}, and the scalar product we have defined 
appears as its only natural lattice regularization. We believe however that this might be too naive. In particular, an arbitrary state of the lattice model expands on an infinite sum involving scaling states (states associated with low energy eigenvalues) as well as non-scaling ones. Depending on the quantities one calculates, these non-scaling states might add up to non-vanishing contributions, even in the limit of a large system.

In the next section we give a strong numerical check that the
lattice scalar products defined above for the Ising chain
and for the dense loop model go over to the continuum limit ones as naively expected, for all quantities we shall be interested in.


\subsection[blabla]{Measure of Affleck-Ludwig boundary entropy \footnote{This section can be skipped on the first reading.}}

For a given CFT, different conformal boundary conditions exist
\cite{CardySurface,CardyVerlinde}.
When one perturbs a conformal boundary condition (CBC) by a relevant operator,
it flows towards another CBC under the RG flow. These CBCs and their flows can be
characterized by their \textit{boundary entropy}
\cite{AffleckLudwig}. A CBC can be encoded by a \textit{boundary
  state} \cite{CardyVerlinde}. The scalar product of a boundary state
$\ket{B}$ with the ground state $\ket{0}$ of the conformal Hamiltonian
on the cylinder $L_0 + \bar{L}_0 -c/12$ gives the boundary entropy
$s_B = - \log \left<B | 0\right>$. These numbers are universal
and have been computed analytically for many CFTs and for many
different CBCs. As a first application of the above considerations on
finite-size scalar products, we are are going to recover them (at least numerically)
from finite-size calculations in our two examples: the Ising chain
and the dense loop model.

\paragraph{Affleck-Ludwig term for the critical Ising chain.}
The Hamiltonian
$H_L$ defined in ($\ref{eq:HamIsing}$) 
acts on the $2^L$-dimensional space of spin configurations
$\ket{s}_L$. The ground state $\ket{0}_L = \sum_{\sigma}
\alpha_{s} \ket{s}_L$ is the eigenstate of $H_L$ with the
largest eigenvalue. The $\alpha_{s}$ are real and
positive. Normalize $\ket{0}_L$ such that $\leftidx{_L}{\left<0 | 0 \right>}{_L} =
\sum_{s} \alpha_{s}^2 =1$.  We consider this system on a
semi-infinite cylinder of circumference $L$. Two different CBC are
obtained by taking the Ising spins on the boundary to be fixed,
$|B\rangle = |+ \cdots + \rangle$, or free, $|B\rangle =
\sum_{s} \ket{s}_L$.  The surface contribution to the free
energy is $f^{B}_s(L)= - \log \left( \leftidx{_L}{\left< B | 0 \right>}{_L}
\right)$.  When $L \rightarrow \infty$ we expect this to scale as
\begin{equation}
 \label{sfactor}
 f^{B}_s(L) \simeq f^{B}_s  L + s_B + \mathcal{O} \left( L^{-1} \right)
\end{equation}
where $f^{B}_s$ is the surface free energy per unit lenght, and $s_B$
is a universal term predicted by CFT. It is well-known
\cite{AffleckLudwig,CardyVerlinde} that
$s_{\rm fixed} = - \log \left( \sqrt{2}/2 \right) \simeq 0.34657$ and
$s_{\rm free} = 0$.

Numerically, we compute $f_s^{B}(L)$ for $L=12,14,16,18$ spins.  Then,
writing
$\displaystyle f_{s}^{B} (L) = \sum_{k= -1}^3 \frac{c_k}{L^{k}}$, $s_B$ should be
identified with $c_0$ according to $(\ref{sfactor})$.
We find $s_{\rm fixed} \simeq 0.3471$ and $s_{\rm free} \simeq 0.00008$.
This shows that the (universal part of the) finite-size scalar products
$\leftidx{_L}{\left<B |0 \right>}{_L}$ can indeed be identified with the CFT result
$\left< B | 0 \right>$. This precise result for the Ising model is not
new, and has deep links with calculations of Renyi entanglement
entropies for critical wave functions \cite{StephanPasquier}.


\paragraph{Affleck-Ludwig term for loop models.}

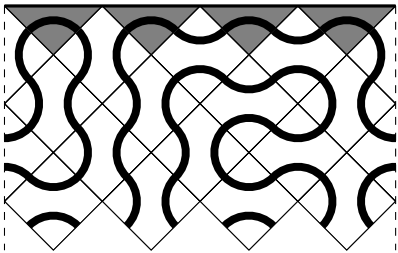
\begin{figure}[h]
	\begin{center}
	\begin{tikzpicture}[scale=1.3]
		\begin{scope}
		\clip (0,0) rectangle (4,2.5);
		\ves{0}{1}
		\vis{1}{1}
		\vis{2}{1}
		\ves{3}{1}
		\ves{4}{1}
		\vis{0}{2}
		\vis{1}{2}
		\ves{2}{2}
		\ves{3}{2}
		\vis{4}{2}

		\ves{-0.5}{0.5}
		\ves{0.5}{0.5}
		\vis{1.5}{0.5}
		\ves{2.5}{0.5}
		\vis{3.5}{0.5}
		\ves{4.5}{0.5}

		\vis{-0.5}{1.5}
		\vis{0.5}{1.5}
		\vis{1.5}{1.5}
		\ves{2.5}{1.5}
		\vis{3.5}{1.5}
		\ves{4.5}{1.5}
	
		\vesg{-0.5}{2.5}
		\vesg{0.5}{2.5}
		\vesg{1.5}{2.5}
		\vesg{2.5}{2.5}
		\vesg{3.5}{2.5}
		\vesg{4.5}{2.5}

		\end{scope}
		\draw[very thick] (0,2.5) -- (4,2.5);
		\draw[dashed] (0,2.5) -- (0,0);
		\draw[dashed] (4,2.5) -- (4,0);
	\end{tikzpicture}
	\end{center}
	\caption{Boundary conditions for dense loops: each loop
          touching the boundary (shaded) is given a weight $n_1$. Bulk
          loops are given a weight $n$. The system is periodic in the
          horizontal direction (dashed lines are identified).}
	\label{fig:blobs}
\end{figure}

To further check this method, we measure the boundary entropy of dense
loops. When $-2 < n \leq 2$ their continuum limit is a CFT with $c = 1
- 6 (\gamma/\pi)^2)/g$, where $g$ and $\gamma$ are parameters related
by $g=1-\gamma/\pi$, and $n = 2 \cos \gamma$.  We use the boundary
condition shown in Fig.~\ref{fig:blobs}: any loop which touches the
boundary (shaded) is given a weight $n_1$ instead of $n$. This is a
CBC for any real $n_1 > 0$, with universal boundary entropy \cite{JS,DJS1}
\begin{equation}
 \label{sloops}
 s_{n,n_1} = - \log \left[ (2g)^{-\frac{1}{4}} \frac{\sin \frac{r \gamma}{g}}
 {\sin r \gamma}
 \left( \frac{\sin \gamma}{\sin \frac{\gamma}{g}} \right)^{\frac{1}{2}}
 \right] \,,
\end{equation}
where $n_1 = \frac{\sin \left((r+1) \gamma\right)}{ \sin( r \gamma)}$ is
parametrized by $r$ \cite{JS,DJS1}.

Let $\ket{0}_L$ be the ground state of $T_L$, normalized such that $\leftidx{_L}{\left<0 |0 \right>}{_L}=1$ for the loop scalar product
$(\ref{loopscalar})$.  The boundary free energy $f^{(n,n_1)}_s (L)$ is
obtained from
$|B\rangle = \ket{ \; \begin{tikzpicture}
\draw[thick] (0,0) arc (-180:0:0.1 and 0.25);
\draw[thick] (0.4,0) arc (-180:0:0.1 and 0.25);
\draw[thick] (0.8,0) arc (-180:0:0.1 and 0.25);
\end{tikzpicture} \; }_L$,
where $\leftidx{_L}{\langle B | 0 \rangle}{_L}$ is computed as in
$(\ref{loopscalar})$, but replacing $n$ by $n_1$ because
the loops touch the boundary (cf.~Fig.~\ref{fig:blobs}).

\begin{figure}[hbpt]
	\centering
	\includegraphics[width=0.75\textwidth]{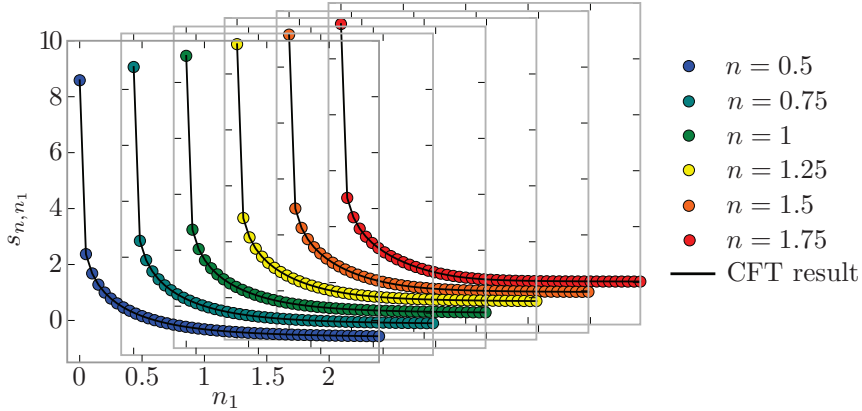}
	\caption{Numerical measures (dots) of
          Affleck-Ludwig boundary entropy for dense loops, as
          functions of $n$ and $n_1$. Solid lines are given by
          (\ref{sloops}).}
	\label{fig:sloops}
\end{figure}

The universal term is computed numerically from $f^{(n,n_1)}_s(L)$ for
$L=14,16,18,20$. The results (see Fig.~\ref{fig:sloops}) are in
excellent agreement with (\ref{sloops}).

\section{The trousers trick}

In the remainder of this paper we will always consider systems with
boundaries, and boundary CFTs. As explained in section 1,
we need to identify a lattice version
of the conformal state $L_{-2} \ket{0}$.
To do this, we first note that $L_{-2}$ generates a
conformal transformation that maps the half-plane with an infinitesimal
vertical slit onto the half-plane itself. This very basic idea is at the
heart of the relation between the operator formalism of CFT and
the celebrated Schramm-Loewner Evolution \cite{BauerBernardCMP,BauerBernard}.
By turning this infinitesimal conformal transformation into a
finite one, we will be able to build its lattice version.

\paragraph{}

\begin{figure}[htbp]
	\centering
	\begin{tikzpicture}[scale=1.3]
		\begin{scope}
			\filldraw[gray] (0,-0.2) rectangle (1.5,2.2);
			\draw[very thick] (0,-0.2) -- (0,2.2);
			\draw[very thick] (1.5,-0.2) -- (1.5,2.2);
			\draw[very thick] (0.75,-0.2) -- (0.75,0.9);
			\draw[<->] (0.1,0) -- (0.65,0);
			\draw[<->] (0.85,0) -- (1.4,0);
			\draw[very thick, dashed] (0,1) -- (1.5,1); 
			\draw (0.75,1.5) node{$\ket{\rm{Trous.}}$}; 
			\draw (0.375,0.3) node{$L/2$};
			\draw (1.125,0.3) node{$L/2$};
		\end{scope}
		
		\draw (1.9,0.9) node{$\sim$};
		\begin{scope}[xshift=3.5cm,yshift=0.2cm]
			\filldraw[gray] (-1.2,0) rectangle (1.2,1.5);
			\draw[very thick] (-1.2,0) -- (1.2,0);
			\draw[very thick] (0,0) -- (0,0.7);
			\draw[very thick, dashed] (-0.75,0) arc (180:0:0.75); 
			\draw (0.4,1.1) node{$\ket{\rm{Trous.}}$}; 
		\end{scope}

		\draw[->] (4.9,0.6) -- (5.8,0.6);
		\draw (5.35,0.9) node{$g_1$};
		\begin{scope}[xshift=7.2cm,yshift=0.2cm]
			\filldraw[gray] (-1.2,0) rectangle (1.2,1.5);
			\draw[very thick] (-1.2,0) -- (1.2,0);
			\draw[very thick, dashed] (-0.75,0) arc (180:0:0.75);
			\draw (0.5,1) node{$\ket{0}$}; 
		\end{scope}
	\end{tikzpicture}
	\caption{The trousers geometry can be developed on the half-plane by the function $g_1(z)=\sqrt{z^2+1}$. This operation is encoded in the state
$\ket{\rm{Trousers}} = {\rm e}^{- \frac{1}{2} L_{-2}}\ket{0}$.}
	\label{fig:conformal}
\end{figure}
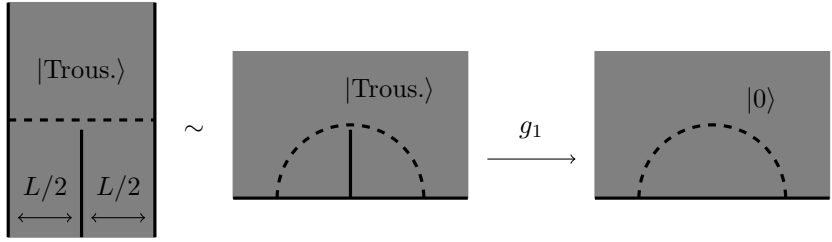

We proceed as in \cite{BauerBernardCMP,BauerBernard}: consider the 
sequence of infinitesimal transformations
$z \mapsto z + {\rm d}g(z)$ where
${\rm d}g(z) = {\rm d}t/(2z)$. This is the infinitesimal
version of the transformation 
\begin{equation}
\label{eq:g_t}
g_t(z) = \sqrt{z^2+t}
\end{equation}
in the sense that $\frac{\partial}{\partial t} g_t (z) = 1/(2 g_t(z))$.
The tranformation ($\ref{eq:g_t}$) maps conformally the half-plane
minus a slit of height $\sqrt{t}$ onto the half-plane (figure $\ref{fig:conformal}$).
Taking $z=g^{-1}_t(w)$, where $w$ is now a point in the upper half-plane,
we have
\begin{equation}
\label{eq:dg_tg_t}
\left( \frac{\partial g_t}{\partial t} \right) \circ g_t^{-1} (w) = \frac{1/2}{w}
\end{equation}
so if $F$ is an arbitrary function defined on the half-plane, it gets
changed under the infinitesimal transformation ${\rm d} g$ by
\begin{equation}
\label{eq:dF}
{\rm d} F =  \frac{{\rm d}t/2}{w} \left(\partial_w F \right) = - \frac{{\rm d}t}{2} \; l_{-2} F
\end{equation}
where $l_{-k} = -w^{-k+1} \partial_w$ is a generator of the Witt algebra.
If $\tilde{G}_g : F \mapsto F \circ g$, we see that equations ($\ref{eq:dg_tg_t}$)
and ($\ref{eq:dF}$) can be written
$\tilde{G}^{-1}_{g_t} \circ \frac{\partial
\tilde{G}_{g_t}}{\partial t} = - (1/2) l_{-2}$  (the order comes
from $\tilde{G}_{f \circ h} = \tilde{G}_h \circ \tilde{G}_f$). We
have thus obtained a differential equation which allows to
compute $\tilde{G}_{g_t}$. This operator maps a function $F$ (defined
on the upper half-plane) onto a function $F \circ g_t$ (defined in
the upper half-plane minus a slit as in Fig. $\ref{fig:conformal}$).
In CFT, we want to act on the Verma modules rather than on simple
functions. Thus we are looking for a \textit{linear} operator $G_{g_t}$
that maps a state $\ket{s}$ of the theory in the upper half-plane
geometry onto $G_{g_t} \ket{s}$, encoding the new
geometry. Turning the Witt generators into the Virasoro ones,
we get
\begin{equation}
G^{-1}_{g_t} . \frac{\partial G_{g_t}}{\partial t} = - \frac{1}{2} L_{-2}
\end{equation}
and of course at $t=0$ the operator is the identity $G_{g_0} = 1$, so
we end up with 
\begin{equation}
\label{eq:expl2}
G_{g_t} = \exp \left( - \frac{t}{2} L_{-2} \right).
\end{equation}

The fact that the action of $L_{-2}$ encodes a small slit
in the original geometry motivates the introduction of the
``trousers'' geometry (Fig. $\ref{fig:conformal}$). Indeed,
it is well-known that a CFT on the infinite strip of width $L$
is equivalent to a CFT in the upper half-plane (the latter can
be mapped on the former by the transformation $z \to w(z) = \frac{L}{\pi} \log z$). Cutting
a small slit in the half-plane is then equivalent to cutting
a slit in the infinite strip, dividing it into two strips of
width $L/2$ up to some height. This is the trousers geometry
(Fig. \ref{fig:conformal}). The ground state of the 
translation operator along the strip $\frac{\pi}{L} \left(L_0 -\frac{c}{24}\right)$ 
is $\ket{0}$. It is different from the ground state
of the system with two legs of width $L/2$, which is
a tensor product $\ket{\rm{Trousers}} \equiv \ket{0}'
\otimes \ket{0}'$ if $\ket{0}'$
is the ground state of width $L/2$. Formula ($\ref{eq:expl2}$)
shows that 
\begin{equation}
\label{eq:trousers}
\ket{\rm{Trousers}} \;=\; e^{-\frac{1}{2}L_{-2}} \ket{0} \;=\;
\ket{0} - \frac{1}{2} L_{-2}\ket{0} + \dots
\end{equation}

\paragraph{}

We get a lattice version of the trousers geometry if we glue together
two ground states of the transfer matrix $T_{L/2}$ or Hamiltonian $H_{L/2}$,
i.e.,
$\ket{\rm{Trousers}}_L \equiv \ket{0}_{L/2} \otimes \ket{0}_{L/2}$ (see Fig. 
$\ref{fig:latticepants}$ for example).

\section{Measure of $b$}

While we are mostly interested in polymers and percolation in this paper, there remains a certain variety of models one can consider. Indeed, one of the difficulties of the field is that geometrical problems are not defined in the usual terms of local degrees of freedom and hamiltonians, and the purported LCFT one is after might well depend on the kind of questions one wants to ask. For percolation for instance, one can decide to focus on the boundaries of clusters, or on the six-vertex model version, or on one of the supersymmetric versions \cite{Gruzberg,ReadSaleur01,ReadSaleur}. We will discuss this point more in the conclusion. For now, we start with the six-vertex model (or equivalently the XXZ spin chain), largely for pedagogical reasons. We then move on to a purely geometrical set-up for the case of polymers.


\subsection{Measure of $b$ in the XXZ spin chain at $q={\rm e}^{i\frac{\pi}{3}}$}

We consider the $U_q(sl_2)$ symmetric XXZ spin chain with
$q={\rm e}^{i \frac{\pi}{3}}$
\cite{PasquierSaleur}. The Hamiltonian with open boundary conditions is
\begin{equation}
\label{XXZH}
 H_L = \sum_{i=1}^{L-1} \left[ \sigma^x_i \sigma_{i+1}^x +
 \sigma^y_i \sigma_{i+1}^y +
 \frac{q+q^{-1}}{2} \sigma^z_i \sigma_{i+1}^z \right]
 + \frac{q-q^{-1}}{2} \left(\sigma_1^z-\sigma_L^z \right) \;.
\end{equation}
Although $H_L$ is not hermitian, its spectrum is real.
When $q={\rm e}^{i \frac{\pi}{3}}$ it has a Jordan cell structure
\cite{PasquierSaleur} for $L \geq 4$.

\paragraph{}
The low-energy spectrum of $H_L$ is described by a CFT with
$c=0$. $H_L$ and the Virasoro generator are related by
$H_L \approx E_0 + \frac{\pi v_F}{L} L_0 + \mathcal{O}\left(\frac{1}{L^2} \right)$
when $L \rightarrow \infty$. Here the Fermi velocity is $v_F=3\sqrt{3}$
\cite{PasquierSaleur}.

\paragraph{Scalar product for the XXZ chain.}
What is the right scalar product for the XXZ chain? We claim that
this is just the Euclidean scalar product in the spin basis,
\textit{treating $q$ as a formal parameter} (i.e., without
complex conjugation). For example, the $q$-singlet state
$\ket{s}_2 \equiv q^{-1/2} \ua\da - q^{1/2} \da\ua$
has a norm squared $\leftidx{_2}{\left< s | s \right>}{_2} = q + q^{-1}$
(if $q$ had been conjugated, one would have found
$\leftidx{_2}{\left<s |s \right>}{_2} = |q| + |q|^{-1}=2$).

There are different ways of seeing that this is the right
scalar product. One simple way is that the XXZ chain can
be mapped on the dense loop model introduced in section $2$,
with a weight $n=q+q^{-1}$ for each closed loop.
This mapping consists in replacing each $q$-singlet by a
half-loop $\; \ldots \otimes \left( q^{-1/2} \ua\da - q^{1/2} \da\ua
\right) \otimes \dots \; \rightarrow \; \dots  \begin{tikzpicture}
\draw[thick] (0.0,0) arc (-180:0:0.1 and 0.25);
\end{tikzpicture}  \dots\;$. Thus the loop scalar product,
as defined in section $2.2$, is the same as the one we
are considering now. In particular, this scalar product
can be negative, and we have already seen that this is
what makes it a good candidate for being a finite-size
version of the Virasoro scalar product in the scaling
limit. Another check that this scalar product is the right
one is that $H_L$ becomes hermitian for this new scalar product.
This is a property that should be expected, because $H_L$ 
is related to $L_0$ in the scaling limit, and $L_0$ is
hermitian for the Virasoro scalar product.

\paragraph{A detailed example: $L=4$.} To explain our strategy,
let us discuss in full detail the case of the lowest interesting
even size. We consider the $S^z=0$ sector only (note that in general
$\left[H_L , S^z \right]=0$). In the basis
$\left\{\ua\ua\da\da, \ua\da\ua\da, \ua\da\da\ua,\da\ua\ua\da ,\da\ua\da\ua, \da\da\ua\ua\right\}$, the Hamiltonian is
$$
H_4 = \left( \begin{array}{cccccc}
\frac{1}{2} + i \sqrt{3} & 2 & 0 & & &  \\ \\
2    &  - \frac{3}{2}+i \sqrt{3} &  2 & 2    \\ \\
0  &   2   &   -\frac{1}{2}  &   0   &  2  \\ \\
 &  2  &  0  &  -\frac{1}{2}   &  2   & 0 \\ \\
  &    &   2  &  2   &  - \frac{3}{2} - i \sqrt{3}  & 2 \\ \\
 & & &  0  &  2  &  \frac{1}{2} - i \sqrt{3}
\end{array} \right)
$$
and it can be put in Jordan form in a suitable basis $\left\{
\ket{0}_4, \ket{1}_4, \ket{2}_4, \ket{3}_4, \ket{\tilde{3}'}_4, \ket{4}_4 \right\}$
$$
H_4 = \left( \begin{array}{cccccc}
- 9/2 &   \\ \\
 &  -2 \sqrt{2} - \frac{1}{2}       \\ \\
  &      &   -\frac{1}{2}  &    \\ \\
 &    &    &  \frac{3}{2}   &    1  \\ \\
  &    &   & 0 &  \frac{3}{2} \\ \\
 & & &    &    &  2 \sqrt{2} - \frac{1}{2}
\end{array} \right)
$$
We give the states $\ket{0}_4$, $\ket{3}_4$ and $\ket{\tilde{3}'}_4$ only (the
other ones are not important in what follows)
$$
\begin{array}{rcl}
\ket{0}_4 &=& \frac{1}{6} \left(-1 + i \sqrt{3}\right) \ua\ua\da\da
	  + \frac{1}{3} \left(2 - i \sqrt{3}\right) \ua\da\ua\da
          - \frac{2}{3}	 \left( \ua\da\da\ua + \da\ua\ua\da \right) \\
& & + \frac{1}{3} \left( 2 + i\sqrt{3}  \right) \da\ua\da\ua
	- \frac{1}{6} \left(1 + i\sqrt{3}\right) \da\da\ua\ua\\
\ket{3}_4 &=& \frac{1}{2} \left(1 + i \sqrt{3}\right) \ua\ua\da\da
	  + \frac{1}{2} \left(-1 - i \sqrt{3}\right) \ua\da\ua\da
          - \left( \ua\da\da\ua + \da\ua\ua\da \right) \\
& & + \frac{1}{2} \left( -1 + i\sqrt{3}  \right) \da\ua\da\ua
	+ \frac{1}{2} \left(1 - i\sqrt{3}\right) \da\da\ua\ua\\
\ket{\tilde{3}'}_4 &=& \frac{1}{4} \left(1 - i \sqrt{3}\right) \ua\ua\da\da
          - \frac{1}{8} \left(1+i 3\sqrt{3}\right) \left( 
		\ua\da\da\ua + \da\ua\ua\da \right) \\
& & + \frac{1}{8} \left( -7 - i\sqrt{3} \right) \da\ua\da\ua
	+ \frac{3}{8} \left(-3 + i\sqrt{3}\right) \da\da\ua\ua\\
\end{array}
$$
With the above scalar product (i.e., without complex conjugation), we see that
$$
\begin{array}{rcl}
	\leftidx{_4}{\left< 0 | 0 \right>}{_4} &=& 1 \\
	\leftidx{_4}{\left< 3 | 3 \right>}{_4} &=& 0 \\
	\leftidx{_4}{\left< 3 | \tilde{3}' \right>}{_4} &=& - \frac{3}{4} .
\end{array}
$$
Note that $\ket{\tilde{3}'}_4$ is only defined up to some
additional term  $\ket{\tilde{3}'}_4 + \lambda \ket{3}_4$. 
The scalar product $\leftidx{_4}{\left< \tilde{3}' | \tilde{3}' \right>}{_4}$
is not invariant under such a transformation, so this is
certainly not a well-defined quantity. On the contrary,
$\leftidx{_4}{\left< 3 | \tilde{3}' \right>}{_4}$ seems well-defined.
However, since $\leftidx{_4}{\left< 3 | 3 \right>}{_4} =0$, we see that there
is still some undetermination: the structure of the Jordan
cell is the same if one rescales $\ket{3}_4 \rightarrow \alpha
\ket{3}_4$ and $\ket{\tilde{3}'}_4 \rightarrow \alpha \ket{\tilde{3}'}_4$,
thus changing the scalar product $\leftidx{_4}{\left< 3| \tilde{3}' \right>}{_4}$.
The whole point of the trousers trick will be to get rid
of this undetermined factor $\alpha$.

To prepare the comparison with CFT, we introduce another
normalization. Let $\ket{\tilde{3}}_4 \equiv \frac{\pi v_F}{4} \ket{\tilde{3}'}_4$,
then in the basis $\left\{ \ket{0}_4, \ket{3}_4 , \ket{\tilde{3}}_4\right\}$
we have
$$
\frac{4}{\pi v_F} \left( H_4 - E_{4,0} \right) = \left(
\begin{array}{ccc}
0 \\
 & \Delta_4  & 1 \\
	& & \Delta_4 
\end{array} \right) 
$$
where $E_{4,0}= -\frac{9}{2}$ is the ground state energy, and
$\Delta_4 = \frac{4 }{\pi v_F} \left( \frac{3}{2} - E_{4,0}\right) = \frac{4 \sqrt{3}}{\pi}$.
The operator $\frac{L}{\pi v_F} \left(H_L-E_{L,0}\right)$ should
be viewed as a lattice version of $L_0$, as follows from the boundary
analog of (\ref{finitesize_cylinder}).

We still have to build the state $\ket{\rm{Trousers}}_4$ for $L=4$.
This is just a tensor product of two times the ground state of
$H_2$, which is a $U_q(sl_2)$ singlet: $\ket{s}_2 \equiv q^{-1/2} \ua\da - q^{1/2} \da\ua$.
$$
\ket{\rm{Trousers}}_4 \; \equiv \; \ket{s}_2 \otimes \ket{s}_2 \;=\;  \left(\frac{1}{2}-i \frac{\sqrt{3}}{2}\right) \ua\da\ua\da
- \left( \ua\da\da\ua + \da\ua\ua\da \right) 
+ \left( \frac{1}{2} + i\frac{\sqrt{3}}{2} \right) \da\ua\da\ua.
$$ 
Note that it is normalized such that $\leftidx{_4}{\left<\rm{Trousers} | 0\right>}{_4}=1$,
in order to match $\langle 0 | 0 \rangle = 1$ when using (\ref{eq:trousers}).

Now we are ready to define a lattice quantity $b(L=4)$, that
is invariant under a global rescaling of the Jordan cell
$\ket{3}_4 \rightarrow \alpha \ket{3}_4$ and $\ket{\tilde{3}}_4
\rightarrow \ket{\tilde{3}}_4$ and that
will later correspond to $b$ in the $L \rightarrow \infty$ limit (see below):
$$
b(4) \;\equiv \; 4 \; \frac{\left( \leftidx{_4}{\left< \rm{Trousers} | \tilde{3} \right>}{_4} \right)^2}{\leftidx{_4}{\left<3 | \tilde{3}\right>}{_4}}.
$$
With the above expressions of the states one gets
$b(4) = - \frac{\sqrt{3} \pi}{4} = -1.3603495\dots \;$.

\paragraph{General strategy.}
We have to restrict to sizes $L$ that are multiples of $4$, because
one needs to build the state $\ket{\rm{Trousers}}$ out of the ground state
of $H_{L/2}$ in the $S^z=0$ sector, so $L/2$ must be even.

If the spectrum of $H_L$ in the $S^z=0$ sector is $E_{L,0} < E_{L,1} < E_{L,2} < \dots$
then $E_{L,3}$ is always twice degenerate (for $L \geq 4$). Let
$\left\{\ket{3}_L,\ket{\tilde{3}}_L\right\}$ be a basis of this equal-energy
subspace, so that the lattice version of $L_0$ reads
\begin{equation}
\frac{L}{\pi v_F} \left( H_L - E_{L,0} \right) = 
\left( \begin{array}{cc}
  \Delta_L & 1 \\
  0 & \Delta_L 
\end{array} \right) \,,
\end{equation}
where $\Delta_L \equiv \frac{L}{\pi v_F} \left(E_{L,3} -
  E_{L,0}\right) \to 2$ as $L \to \infty$.

Again, there is an
undetermined overall normalization because $\leftidx{_L}{\left< 3|3
\right>}{_L}=0$ (with the same scalar product as above). 
The trousers trick is used again to
define a normalization-independent lattice quantity $b(L)$ as in the
case $L=4$. We normalize $\ket{0}_L$ and $\ket{\rm{Trousers}}_L$ such
that $\leftidx{_L}{\left< \rm{Trousers}|0 \right>}{_L} \;=\; \leftidx{_L}{\left< 0|0 \right>}{_L} \;= \;1$. Then define
\begin{equation}
b(L) \equiv 4 \frac{\left( \leftidx{_L}{\left< \rm{Trousers} | \tilde{3} \right>}{_L} \right)^2}{\leftidx{_L}{\left<3 | \tilde{3}\right>}{_L}}.
\end{equation}
When $L \to \infty$, we expect $\ket{3}_L \sim \alpha L_{-2} \ket{0}$
and $\ket{\tilde{3}}_L \sim \alpha \Phi_{1,5} \ket{0}$. Using relation
($\ref{eq:trousers}$) we get
\begin{equation}
\label{eq:bXXZ}
b(L) \quad \underset{L \rightarrow \infty}{\longrightarrow} \quad
4 \;\frac{\left(\alpha \left< \rm{Trousers} | \Phi_{1,5}
| 0 \right> \right)^2}{\alpha^2 \left<0 | L_2 \Phi_{1,5}|0 \right>}
\; = \; \left<0 | L_2 \Phi_{1,5} |0 \right> \; = \; \langle T | t \rangle \; = \; b.
\end{equation}
Numerical results are given below (Tab. $\ref{tab:numeric}$).


\subsection{Measure of $b$ in the dilute polymers model}

We consider a dilute polymers model on the honeycomb lattice.
It is defined by the tranfer matrix (in the picture $L=6$)
$$\begin{tikzpicture}[scale=0.40]
	\draw (-1.7,0) node {$T_L \; =$};
	\draw (0,0) -- (1,-0.577) -- (4,1.1547) -- (6,0) -- (6,1.1547) -- (3,-0.577) -- (0,1.1547) -- cycle;
	\draw (0,0) -- (2,1.1547) -- (5,-0.577) -- (6,0);
\end{tikzpicture}$$
where each losange is a sum over eight configurations with
weight $x$ per monomer:
$$
\begin{tikzpicture}
	\begin{scope}[scale=0.4]
		\draw[clip] (0,0) -- (1,-0.577) -- (2,0) -- (1,0.577) -- cycle;
		\draw[gray,thick] (0.333,0.577) -- (0.667,0.) -- (0.333,-0.577);
		\draw[gray,thick] (1.667,0.577) -- (1.333,0.) -- (1.667,-0.577);
		\draw[gray,thick] (0.667,0.) -- (1.333,0.);
	\end{scope}
	\draw[scale=0.4] (1,-1.3) node{$1$};

	\begin{scope}[scale=0.4,xshift=2.5cm]
		\draw[clip] (0,0) -- (1,-0.577) -- (2,0) -- (1,0.577) -- cycle;
		\draw[gray,thick] (0.333,0.577) -- (0.667,0.) -- (0.333,-0.577);
		\draw[gray,thick] (1.667,0.577) -- (1.333,0.) -- (1.667,-0.577);
		\draw[gray,thick] (0.667,0.) -- (1.333,0.);
		\draw[line width=2pt] (0.333,0.577) -- (0.667,0.) -- (0.333,-0.577);
	\end{scope}
	\draw[scale=0.4,xshift=2.5cm] (1,-1.35) node{$x$};

	\begin{scope}[scale=0.4,xshift=5cm]
		\draw[clip] (0,0) -- (1,-0.577) -- (2,0) -- (1,0.577) -- cycle;
		\draw[gray,thick] (0.333,0.577) -- (0.667,0.) -- (0.333,-0.577);
		\draw[gray,thick] (1.667,0.577) -- (1.333,0.) -- (1.667,-0.577);
		\draw[gray,thick] (0.667,0.) -- (1.333,0.);
		\draw[line width=2pt] (1.667,0.577) -- (1.333,0.) -- (1.667,-0.577);
	\end{scope}
	\draw[scale=0.4,xshift=5cm] (1,-1.35) node{$x$};
	
	\begin{scope}[scale=0.4,xshift=7.5cm]
		\draw[clip] (0,0) -- (1,-0.577) -- (2,0) -- (1,0.577) -- cycle;
		\draw[gray,thick] (0.333,0.577) -- (0.667,0.) -- (0.333,-0.577);
		\draw[gray,thick] (1.667,0.577) -- (1.333,0.) -- (1.667,-0.577);
		\draw[gray,thick] (0.667,0.) -- (1.333,0.);
		\draw[line width=2pt] (1.667,0.577) -- (1.333,0.) -- (1.667,-0.577);
		\draw[line width=2pt] (0.333,0.577) -- (0.667,0.) -- (0.333,-0.577);
	\end{scope}
	\draw[scale=0.4,xshift=7.5cm] (1,-1.2) node{$x^2$};
	
	\begin{scope}[scale=0.4,xshift=10cm]
		\draw[clip] (0,0) -- (1,-0.577) -- (2,0) -- (1,0.577) -- cycle;
		\draw[gray,thick] (0.333,0.577) -- (0.667,0.) -- (0.333,-0.577);
		\draw[gray,thick] (1.667,0.577) -- (1.333,0.) -- (1.667,-0.577);
		\draw[gray,thick] (0.667,0.) -- (1.333,0.);
		\draw[line width=2pt] (1.667,0.577) -- (1.333,0.) -- (0.667,0) -- (0.333,0.577);
	\end{scope}
	\draw[scale=0.4,xshift=10cm] (1,-1.2) node{$x^2$};
	
	\begin{scope}[scale=0.4,xshift=12.5cm]
		\draw[clip] (0,0) -- (1,-0.577) -- (2,0) -- (1,0.577) -- cycle;
		\draw[gray,thick] (0.333,0.577) -- (0.667,0.) -- (0.333,-0.577);
		\draw[gray,thick] (1.667,0.577) -- (1.333,0.) -- (1.667,-0.577);
		\draw[gray,thick] (0.667,0.) -- (1.333,0.);
		\draw[line width=2pt] (1.667,-0.577) -- (1.333,0.) -- (0.667,0) -- (0.333,-0.577);
	\end{scope}
	\draw[scale=0.4,xshift=12.5cm] (1,-1.2) node{$x^2$};
	
	\begin{scope}[scale=0.4,xshift=15cm]
		\draw[clip] (0,0) -- (1,-0.577) -- (2,0) -- (1,0.577) -- cycle;
		\draw[gray,thick] (0.333,0.577) -- (0.667,0.) -- (0.333,-0.577);
		\draw[gray,thick] (1.667,0.577) -- (1.333,0.) -- (1.667,-0.577);
		\draw[gray,thick] (0.667,0.) -- (1.333,0.);
		\draw[line width=2pt] (1.667,0.577) -- (1.333,0.) -- (0.667,0) -- (0.333,-0.577);
	\end{scope}
	\draw[scale=0.4,xshift=15cm] (1,-1.2) node{$x^2$};
	
	\begin{scope}[scale=0.4,xshift=17.5cm]
		\draw[clip] (0,0) -- (1,-0.577) -- (2,0) -- (1,0.577) -- cycle;
		\draw[gray,thick] (0.333,0.577) -- (0.667,0.) -- (0.333,-0.577);
		\draw[gray,thick] (1.667,0.577) -- (1.333,0.) -- (1.667,-0.577);
		\draw[gray,thick] (0.667,0.) -- (1.333,0.);
		\draw[line width=2pt] (1.667,-0.577) -- (1.333,0.) -- (0.667,0) -- (0.333,0.577);
	\end{scope}
	\draw[scale=0.4,xshift=17.5cm] (1,-1.2) node{$x^2$};
\end{tikzpicture}
$$
The boundary triangles can appear with or without
monomers, with weights $x$ or $1$. There are no closed
loops in this model (i.e., the loop weight is $n=0$).
This model is the $n \rightarrow 0$ limit of the $O(n)$
model. Its critical point \cite{Nienhuis82}
is $x = (2 + \sqrt{2})^{-1/2}$, and it corresponds in the
scaling limit to dilute polymers (or self-avoiding walks).

The transfer matrix $T_L$ acts on configuration states
that contain half-loops and empty sites (marked with dots), such as 
$\ket{\; \begin{tikzpicture}
\draw[thick] (0.4,0) arc (-180:0:0.3 and 0.25);
\draw[thick] (0.,-0.) circle (0.01);
\draw[thick] (0.2,-0.) circle (0.01);
\draw[thick] (0.6,-0.) circle (0.01);
\draw[thick] (0.8,-0.) circle (0.01);
\end{tikzpicture}\;}$
or
$\ket{\; \begin{tikzpicture}
\draw[thick] (0,0) arc (-180:0:0.4 and 0.25);
\draw[thick] (0.4,0) arc (-180:0:0.1 and 0.15);
\draw[thick] (0.2,-0.) circle (0.01);
\draw[thick] (1,-0.) circle (0.01);
\end{tikzpicture}\;}$ for $L=6$.
We also have to encode the fact that polymers
are already in the system when one applies
$T_L$ the first time, or in other words that
they are ``connected to the infinite past''.
We thus introduce strings (unpaired points
interpreted as loop segments connecting to
the infinite past) like for example
$\ket{\; \begin{tikzpicture}
\draw[thick] (0,0) arc (-180:0:0.1 and 0.25);
\draw[thick] (0.4,-0.) circle (0.01);
\draw[thick] (0.8,-0.) circle (0.01);
\draw[thick] (0.6,-0.25) -- ++(0,0.25);
\draw[thick] (1,-0.25) -- ++(0,0.25);
\end{tikzpicture}\;}$ or 
$\ket{\; \begin{tikzpicture}
\draw[thick] (0.,-0.) circle (0.01);
\draw[thick] (0.4,-0.25) -- ++(0,0.25);
\draw[thick] (0.6,-0.25) -- ++(0,0.25);
\draw[thick] (1,-0.25) -- ++(0,0.25);
\draw[thick] (0.2,-0.) circle (0.01);
\draw[thick] (0.8,-0.) circle (0.01);
\end{tikzpicture}\;}$
with two and three polymers coming from
infinity respectively.
The action of $T_L$ can lower the number of
strings: it can turn out that two polymers
coming from infinity were actually two pieces
of the same polymer connected to infinity.
But iterations of the transfer matrix cannot
create new polymers connected to the past.
This ``irreversibility'' is the origin of
the Jordan cells of $T_L$ in this model. 
To illustrate this, consider the case of
$L=2$ sites. One has 
$T_2 = \begin{tikzpicture}[scale=0.2]
	\draw (0,0) -- (1,-0.577) -- (2,0) -- (0,1.1547) -- cycle;
	\draw (0,0) -- (2,1.1547) -- (2,0);
\end{tikzpicture}$.
In the basis
$\left\{ \ket{\; \begin{tikzpicture}
\draw[thick] (0,0) circle (0.01);
\draw[thick] (0.2,0) circle (0.01);
\draw[white] (0.2,-0.25) circle (0.001);
\end{tikzpicture}\;} , \ket{ \; \begin{tikzpicture}
\draw[thick] (0,0) arc (-180:0:0.1 and 0.25);
\end{tikzpicture}\;} , \ket{ \; \begin{tikzpicture}
\draw[thick] (0,0) -- (0,-0.25);
\draw[thick] (0.2,0) -- (0.2,-0.25);
\end{tikzpicture}\;} \right\}$
it reads
\begin{equation}
\label{eq:T2}
 T_2 = \left( \begin{array}{ccc} 1 & 0 & x^2 \\
	 			x^4 & x^4 & 0  \\
				0  & 0 & x^4  \end{array}  \right).
\end{equation}
$T_2$ has a rank-two Jordan cell for the eigenvalue $x^4$.  The same
structure (of course for different eigenvalues) is found
in $T_L$ for higher $L$.

\paragraph{Scalar product.} We proceed as in section $2.2$
to establish which scalar product is relevant here. The scalar
product has to be compatible with the action of the transfer
matrix. For the basis states with only empty sites and half-loops,
the scalar product is obvious. First, it has to be zero
if the empty sites are not the same in the two states. Then
one just uses the loop scalar product (section $2.2$), which
is actually zero as soon as there is a closed loop
(recall that $n=0$). For example 
$\left<  \; \begin{tikzpicture}
\draw[thick] (0,0) arc (-180:0:0.1 and 0.25);
\draw[thick] (0.6,-0.) circle (0.01);
\draw[thick] (0.4,-0.) circle (0.01);
\end{tikzpicture}\; |  \; \begin{tikzpicture}
\draw[thick] (0,-0.) circle (0.01);
\draw[thick] (0.2,0) arc (-180:0:0.2 and 0.25);
\draw[thick] (0.4,-0.) circle (0.01);
\end{tikzpicture}\;\right> \;= \;0\;$ because the empty sites
are not the same, and  
$\left<  \; \begin{tikzpicture}
\draw[thick] (0,-0.) circle (0.01);
\draw[thick] (0.2,0) arc (-180:0:0.2 and 0.25);
\draw[thick] (0.4,-0.) circle (0.01);
\end{tikzpicture}\; |  \; \begin{tikzpicture}
\draw[thick] (0,-0.) circle (0.01);
\draw[thick] (0.2,0) arc (-180:0:0.2 and 0.25);
\draw[thick] (0.4,-0.) circle (0.01);
\end{tikzpicture}\;\right> \;= \;0\;$ because there is a
closed loop. Actually, so far the only basis state that
contributes to non-zero scalar products is
$\left<  \; \begin{tikzpicture}
\draw[thick] (0,-0.) circle (0.01);
\draw[thick,white] (0,-0.25) circle (0.01);
\draw[thick] (0.2,-0.) circle (0.01);
\draw[thick] (0.6,-0.) circle (0.01);
\draw[thick] (0.4,-0.) circle (0.01);
\end{tikzpicture}\; |  \; \begin{tikzpicture}
\draw[thick] (0,-0.) circle (0.01);
\draw[thick,white] (0,-0.25) circle (0.01);
\draw[thick] (0.2,-0.) circle (0.01);
\draw[thick] (0.6,-0.) circle (0.01);
\draw[thick] (0.4,-0.) circle (0.01);
\end{tikzpicture}\;\right> \;= \;1$.

What about the strings connected to the infinite past?
Since two strings can be connected by the transfer
matrix, this should also be allowed by the scalar product.
Thus we define the scalar product as $1$ if
one contracts a pair of strings against a
half-loop, like 
$\left< \; \begin{tikzpicture}
\draw[thick] (0,-0.) circle (0.01);
\draw[thick] (0.2,0) arc (-180:0:0.2 and 0.25);
\draw[thick] (0.4,-0.) circle (0.01);
\end{tikzpicture}\; |  \; \begin{tikzpicture}
\draw[thick] (0,-0.) circle (0.01);
\draw[thick] (0.4,-0.) circle (0.01);
\draw[thick] (0.2,-0.25) -- ++(0,0.25);
\draw[thick] (0.6,-0.25) -- ++(0,0.25);
\end{tikzpicture}\;\right> \; = \; 1$.

\paragraph{Detailed calculation of the lattice $b(L)$
for $L=2$.} Like for the XXZ chain, we give here
a concrete example of our strategy for the smallest
interesting size. We start from the transfer matrix
($\ref{eq:T2}$) which can be put in Jordan form
in the basis $\left\{ \ket{0}_2, \ket{1}_2,\ket{\tilde{1}'}_2\right\}$
$$
T_2 = \left( \begin{array}{ccc} 1 &  &  \\
	 			 & x^4 & 1  \\
				  &  & x^4  \end{array}  \right) \,,
$$
where
$$
\begin{array}{rcl}
\ket{0}_2 &=& \ket{\; \begin{tikzpicture}
\draw[thick] (0,0) circle (0.01);
\draw[thick] (0.2,0) circle (0.01);
\draw[white] (0.2,-0.25) circle (0.001);
\end{tikzpicture}\;}  + \frac{x^4}{1-x^4}
 \ket{ \; \begin{tikzpicture}
\draw[thick] (0,0) arc (-180:0:0.1 and 0.25);
\end{tikzpicture}\;}  \\

\ket{1}_2 &=& - \frac{x^6}{1-x^4} \ket{ \; \begin{tikzpicture}
\draw[thick] (0,0) arc (-180:0:0.1 and 0.25);
\end{tikzpicture}\;}  \\

\ket{\tilde{1}'}_2 &=& \ket{ \; \begin{tikzpicture}
\draw[thick] (0,0) -- (0,-0.25);
\draw[thick] (0.2,0) -- (0.2,-0.25);
\end{tikzpicture}\;} - \frac{x^2}{1-x^4}
\ket{\; \begin{tikzpicture}
\draw[thick] (0,0) circle (0.01);
\draw[thick] (0.2,0) circle (0.01);
\draw[white] (0.2,-0.25) circle (0.001);
\end{tikzpicture}\;} .
\end{array}
$$
The state $\ket{0}_2$ is normalized so that
$\left< \;\begin{tikzpicture}
\draw[thick] (0,0) circle (0.01);
\draw[thick] (0.2,0) circle (0.01);
\draw[white] (0.2,-0.2) circle (0.001);
\end{tikzpicture}\; |0 \right>_2 = 1$. We also
need the left eigenstates of $T_2$ (see section $2$)
$$
\begin{array}{rcl}
\leftidx{_2}{\bra{0}}{} &=& \bra{\; \begin{tikzpicture}
\draw[thick] (0,0) circle (0.01);
\draw[thick] (0.2,0) circle (0.01);
\draw[white] (0.2,0.25) circle (0.001);
\end{tikzpicture}\;}  + \frac{x^2}{1-x^4}
 \bra{ \; \begin{tikzpicture}
\draw[thick] (0,0) arc (180:0:0.1 and 0.25);
\end{tikzpicture}\;}  \\

\leftidx{_2}{\bra{1}}{} &=& - \frac{x^6}{1-x^4} \bra{ \; \begin{tikzpicture}
\draw[thick] (0,0) arc (180:0:0.1 and 0.25);
\end{tikzpicture}\;}  \\

\leftidx{_2}{\bra{\tilde{1}'}} &=& \bra{ \; \begin{tikzpicture}
\draw[thick] (0,0) -- (0,-0.25);
\draw[thick] (0.2,0) -- (0.2,-0.25);
\end{tikzpicture}\;} - \frac{x^4}{1-x^4}
\bra{\; \begin{tikzpicture}
\draw[thick] (0,0) circle (0.01);
\draw[thick] (0.2,0) circle (0.01);
\draw[white] (0.2,0.25) circle (0.001);
\end{tikzpicture}\;} .
\end{array}
$$
Note that this is not a simple transposition
of the right Jordan basis, and that here the strings
represent polymers connected to the infinite
future (not past). There are two independent undetermined
overall factors in the normalizations of the
cells $\left\{ \ket{1}_2, \ket{\tilde{1}'}_2 \right\}$
and $\left\{ \leftidx{_2}{\bra{1}}{}, \leftidx{_2}{\bra{\tilde{1}'}}{} \right\}$.
They will be fixed later using the trousers trick.

Like in the XXZ chain, we introduce another normalization
of the Jordan cell to relate our results to CFT. Let
$\Delta_2 \equiv - \frac{\sqrt{3}}{2} \frac{2}{\pi} \log
\left( x^4 \right)$, and $\ket{\tilde{1}}_2 \equiv
- \frac{2}{\sqrt{3}} \frac{\pi}{2} x^4 \ket{\tilde{1}'}_2$. Then in the
basis $\left\{ \ket{1}_2 , \ket{\tilde{1}}_2 \right\}$
$$
T_2 = \exp \left[ -\frac{2}{\sqrt{3}} \frac{\pi}{2}
\left( \begin{array}{cc} \Delta_2  & 1  \\
	 			 0  & \Delta_2  \end{array}  \right) \right].
$$
The same normalization will be used for the left Jordan cell:
$\leftidx{_2}{\bra{\tilde{1}}}{} \equiv
- \frac{2}{\sqrt{3}} \frac{\pi}{2} x^4 \; \leftidx{_2}{\bra{\tilde{1}'}}{}$.
Now we introduce the trousers state. It turns out that it is
trivial for $L=2$, because the ground state of the transfer
matrix on $L/2$ sites, i.e. $T_1$, is trivial. Then we have
simply $\ket{\rm{Trousers}}_2 = \ket{\;\begin{tikzpicture}
\draw[thick] (0,0) circle (0.01);
\draw[thick] (0.2,0) circle (0.01);
\draw[white] (0.2,-0.2) circle (0.001);
\end{tikzpicture}\;}$. It is normalized
so that $\left< \;\begin{tikzpicture}
\draw[thick] (0,0) circle (0.01);
\draw[thick] (0.2,0) circle (0.01);
\draw[white] (0.2,0.2) circle (0.001);
\end{tikzpicture}\; | \rm{Trousers} \right>_2 \; =\; 1$. 
The same is true of course
for the left trousers state  $\; \leftidx{_2}{\bra{\rm{Trousers}}}{}
\; = \; \bra{\;\begin{tikzpicture}
\draw[thick] (0,0) circle (0.01);
\draw[thick] (0.2,0) circle (0.01);
\draw[white] (0.2,0.2) circle (0.001);
\end{tikzpicture}\;}$. Now let us define
$$
b(2) \; \equiv \; 4\; \frac{\left(\leftidx{_2}{\left< \rm{Trousers} |
\tilde{1}\right>}{_2} \right) \left(\leftidx{_2}{\left< \tilde{1} |
\rm{Trousers} \right>}{_2} \right)}{\leftidx{_2}{\left< 1 | \tilde{1} \right>}{_2}}. 
$$
This quantity is invariant under a rescaling of the
right Jordan cell $\ket{1}_2 \rightarrow \alpha \ket{1}_2$,  
$\ket{\tilde{1}}_2 \rightarrow \alpha \ket{\tilde{1}}_2$. It
is also invariant under a rescaling of the left one
$\; \leftidx{_2}{\bra{1}}{} \rightarrow \beta \; \leftidx{_2}{\bra{1}}{}$,  
$\; \leftidx{_2}{\bra{\tilde{1}}}{} \rightarrow \beta \; \leftidx{_2}{\bra{\tilde{1}}}{}$,
so it does not depend on the particular normalizations of
the states we have chosen. With the above
formulas,
one can easily check that $b(2) = \frac{4 \pi}{\sqrt{3}}
\frac{ x^4}{1-x^4} = 0.680801\dots$
(recall that $x = (2 + \sqrt{2})^{-1/2}$).

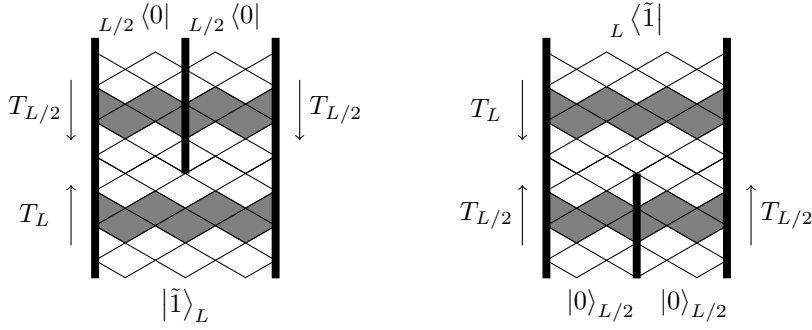
\begin{figure}[htbp]
\centering
	\begin{tikzpicture}

	\begin{scope}[scale=0.40]
		\begin{scope}
		\draw (0,0) -- (1,-0.577) -- (4,1.1547) -- (6,0) -- (6,1.1547) -- (3,-0.577) -- (0,1.1547) -- cycle;
		\draw (0,0) -- (2,1.1547) -- (5,-0.577) -- (6,0);
		\end{scope}
		\begin{scope}[yshift=2.3094cm]
		\draw (0,0) -- (1,-0.577) -- (4,1.1547) -- (6,0) -- (6,1.1547) -- (3,-0.577) -- (0,1.1547) -- cycle;
		\draw (0,0) -- (2,1.1547) -- (5,-0.577) -- (6,0);
		\end{scope}
		\begin{scope}[yshift=1.1547cm]
		\filldraw[gray] (0,0) -- (1,0.577) -- (0,1.1547) -- cycle;
		\filldraw[gray] (4,0) -- (5,-0.577) -- (6,0) -- (5,0.577) -- cycle;
		\filldraw[gray] (2,0) -- (3,-0.577) -- (4,0) -- (3,0.577) -- cycle;
		\filldraw[gray] (0,0) -- (1,-0.577) -- (2,0) -- (1,0.577) -- cycle;
		\filldraw[gray] (1,0.577) -- (2,1.1547) -- (3,0.577) -- (2,0) -- cycle;
		\filldraw[gray] (3,0.577) -- (4,1.1547) -- (5,0.577) -- (4,0) -- cycle;
		\filldraw[gray] (6,0) -- (5,0.577) -- (6,1.1547) -- cycle;
		\draw (0,0) -- (1,-0.577) -- (4,1.1547) -- (6,0) -- (6,1.1547) -- (3,-0.577) -- (0,1.1547) -- cycle;
		\draw (0,0) -- (2,1.1547) -- (5,-0.577) -- (6,0);
		\end{scope}
		\begin{scope}[yshift=3.4641cm]

		\draw (0,0) -- (1,-0.577) -- (4,1.1547) -- (6,0) -- (6,1.1547) -- (3,-0.577) -- (0,1.1547) -- cycle;
		\draw (0,0) -- (2,1.1547) -- (5,-0.577) -- (6,0);
		\end{scope}
		\begin{scope}[yshift=5.7735cm]
		\draw (0,0) -- (1,-0.577) -- (4,1.1547) -- (6,0) -- (6,1.1547) -- (3,-0.577) -- (0,1.1547) -- cycle;
		\draw (0,0) -- (2,1.1547) -- (5,-0.577) -- (6,0);
		\end{scope}
		\begin{scope}[yshift=4.6188cm]
		\filldraw[gray] (0,0) -- (1,0.577) -- (0,1.1547) -- cycle;
		\filldraw[gray] (4,0) -- (5,-0.577) -- (6,0) -- (5,0.577) -- cycle;
		\filldraw[gray] (2,0) -- (3,-0.577) -- (4,0) -- (3,0.577) -- cycle;
		\filldraw[gray] (0,0) -- (1,-0.577) -- (2,0) -- (1,0.577) -- cycle;
		\filldraw[gray] (1,0.577) -- (2,1.1547) -- (3,0.577) -- (2,0) -- cycle;
		\filldraw[gray] (3,0.577) -- (4,1.1547) -- (5,0.577) -- (4,0) -- cycle;
		\filldraw[gray] (6,0) -- (5,0.577) -- (6,1.1547) -- cycle;
		\draw (0,0) -- (1,-0.577) -- (4,1.1547) -- (6,0) -- (6,1.1547) -- (3,-0.577) -- (0,1.1547) -- cycle;
		\draw (0,0) -- (2,1.1547) -- (5,-0.577) -- (6,0);
		\end{scope}

		\draw[line width=3pt] (0,-0.6) -- (0,7.4); 
		\draw[line width=3pt] (6,-0.6) -- (6,7.4); 
		\draw[line width=3pt] (3,7.4) -- (3,2.885);

		\draw[->] (-0.8,0.5) -- (-0.8,2.5);
		\draw (-2,1.5) node{$T_L$}; 
		\draw[->] (-0.8,6) -- (-0.8,4);
		\draw[->] (6.8,6) -- (6.8,4);
		\draw (-2,5) node{$T_{L/2}$}; 
		\draw (8,5) node{$T_{L/2}$};
		\draw (3,-1.5) node{$\ket{\tilde{1}}_L$};  
		\draw (1.3,8) node{$\leftidx{_{L/2}}{\bra{0}}{}$};  
		\draw (4.3,8) node{$\leftidx{_{L/2}}{\bra{0}}{}$};  
	\end{scope}

	\begin{scope}[xshift=6cm,scale=0.40]
		\begin{scope}
		\draw (0,0) -- (1,-0.577) -- (4,1.1547) -- (6,0) -- (6,1.1547) -- (3,-0.577) -- (0,1.1547) -- cycle;
		\draw (0,0) -- (2,1.1547) -- (5,-0.577) -- (6,0);
		\end{scope}
		\begin{scope}[yshift=2.3094cm]
		\draw (0,0) -- (1,-0.577) -- (4,1.1547) -- (6,0) -- (6,1.1547) -- (3,-0.577) -- (0,1.1547) -- cycle;
		\draw (0,0) -- (2,1.1547) -- (5,-0.577) -- (6,0);
		\end{scope}
		\begin{scope}[yshift=1.1547cm]
		\filldraw[gray] (0,0) -- (1,0.577) -- (0,1.1547) -- cycle;
		\filldraw[gray] (4,0) -- (5,-0.577) -- (6,0) -- (5,0.577) -- cycle;
		\filldraw[gray] (2,0) -- (3,-0.577) -- (4,0) -- (3,0.577) -- cycle;
		\filldraw[gray] (0,0) -- (1,-0.577) -- (2,0) -- (1,0.577) -- cycle;
		\filldraw[gray] (1,0.577) -- (2,1.1547) -- (3,0.577) -- (2,0) -- cycle;
		\filldraw[gray] (3,0.577) -- (4,1.1547) -- (5,0.577) -- (4,0) -- cycle;
		\filldraw[gray] (6,0) -- (5,0.577) -- (6,1.1547) -- cycle;
		\draw (0,0) -- (1,-0.577) -- (4,1.1547) -- (6,0) -- (6,1.1547) -- (3,-0.577) -- (0,1.1547) -- cycle;
		\draw (0,0) -- (2,1.1547) -- (5,-0.577) -- (6,0);
		\end{scope}
		\begin{scope}[yshift=3.4641cm]

		\draw (0,0) -- (1,-0.577) -- (4,1.1547) -- (6,0) -- (6,1.1547) -- (3,-0.577) -- (0,1.1547) -- cycle;
		\draw (0,0) -- (2,1.1547) -- (5,-0.577) -- (6,0);
		\end{scope}
		\begin{scope}[yshift=5.7735cm]
		\draw (0,0) -- (1,-0.577) -- (4,1.1547) -- (6,0) -- (6,1.1547) -- (3,-0.577) -- (0,1.1547) -- cycle;
		\draw (0,0) -- (2,1.1547) -- (5,-0.577) -- (6,0);
		\end{scope}
		\begin{scope}[yshift=4.6188cm]
		\filldraw[gray] (0,0) -- (1,0.577) -- (0,1.1547) -- cycle;
		\filldraw[gray] (4,0) -- (5,-0.577) -- (6,0) -- (5,0.577) -- cycle;
		\filldraw[gray] (2,0) -- (3,-0.577) -- (4,0) -- (3,0.577) -- cycle;
		\filldraw[gray] (0,0) -- (1,-0.577) -- (2,0) -- (1,0.577) -- cycle;
		\filldraw[gray] (1,0.577) -- (2,1.1547) -- (3,0.577) -- (2,0) -- cycle;
		\filldraw[gray] (3,0.577) -- (4,1.1547) -- (5,0.577) -- (4,0) -- cycle;
		\filldraw[gray] (6,0) -- (5,0.577) -- (6,1.1547) -- cycle;
		\draw (0,0) -- (1,-0.577) -- (4,1.1547) -- (6,0) -- (6,1.1547) -- (3,-0.577) -- (0,1.1547) -- cycle;
		\draw (0,0) -- (2,1.1547) -- (5,-0.577) -- (6,0);
		\end{scope}

		\draw[line width=3pt] (0,-0.6) -- (0,7.4); 
		\draw[line width=3pt] (6,-0.6) -- (6,7.4); 
		\draw[line width=3pt] (3,-0.6) -- (3,2.885);

		\draw[->] (-0.8,0.5) -- (-0.8,2.5);
		\draw (-2,1.5) node{$T_{L/2}$}; 
		\draw[->] (-0.8,6) -- (-0.8,4);
		\draw[->] (6.8,0.5) -- (6.8,2.5);
		\draw (-2,5) node{$T_L$}; 
		\draw (8,1.5) node{$T_{L/2}$};
		\draw (3,8) node{$\leftidx{_L}{\bra{\tilde{1}}}{}$};  
		\draw (1.8,-1.5) node{$\ket{0}_{L/2}$};  
		\draw (4.9,-1.5) node{$\ket{0}_{L/2}$};  
	\end{scope}

	\end{tikzpicture}
\caption{The different lattice geometries used in the computation of $b(L)$
for polymers. The first one (left) is used represents the term $_L \left<
\rm{Trousers} | \tilde{1} \right>_L$, the other one (right) the term
$_L \left< \tilde{1}| \rm{Trousers} \right>_L$.}
\label{fig:latticepants}
\end{figure}

\paragraph{General strategy.} Dilute polymers are described by a CFT with $c=0$
\cite{LoopReview}. $T_L$ is related to the (continuum)
translation operator along the strip $(\pi / L) L_0$ when $L
\rightarrow \infty$ as follows:
\begin{equation}
T_L \simeq \lambda_{L,0} \left[ {\rm e}^{- \frac{2}{\sqrt{3}} \frac{\pi}{L} L_0}
 + \mathcal{O} \left( \frac{1}{L^2} \right) \right] \,,
\end{equation}
where $\lambda_{L,0}>\lambda_{L,1}> \lambda_{L,2} > \ldots$ are the
eigenvalues of $T_L$, and the factor $\sqrt{3}/2$ is due to the honeycomb
lattice. One finds that $\lambda_{L,1}$ is always
degenerate twice. When $L \to \infty$ the first three eigenstates of
$T_L$ should be identified with the CFT states $\ket{0}, L_{-2}\ket{0}$
and $\Phi_{3,1} \ket{0}$ of lowest conformal dimensions ($0$, $2$ and
$2$).  The operator $\Phi_{3,1}$ inserts two strings at the infinite
past \cite{LoopReview}.  One can check numerically that the lattice version
of the conformal eigenvalue $\Delta_L \equiv - \frac{\sqrt{3}}{2}
\frac{L}{\pi} \log \left( \lambda_{L,1}/\lambda_{L,0}\right) $
converges to $2$ when $L \to \infty$.

In the two-dimensional subspace corresponding to $\lambda_{L,1}$, one
can define $\left\{ \ket{1}_L, \ket{\tilde{1}}_L \right\}$ such that in this
basis
$$
T_L =  \lambda_{L,0} \; \exp \left[ - \frac{2}{\sqrt{3}} \frac{\pi}{L}
\left( \begin{array}{cc}   \Delta_L & 1  \\ 0  & \Delta_L \end{array} \right)\right].
$$
This defines a lattice version of the conformal states $L_{-2} \ket{0}
\simeq \alpha \ket{1}_L$ and $\Phi_{3,1} \ket{0} \simeq  \alpha \ket{2}_L$
up to some overall normalization constant $\alpha$. We can do the
same thing for the left action of $T_L$, thus defining
the left Jordan cell $\left\{\leftidx{_L}{\bra{1}}{}, \; \leftidx{_L}{\bra{\tilde{1}}}{} \right\}$,
which should be viewed as 
$\bra{0} L_{2} \simeq \beta \; \leftidx{_L}{\bra{1}}{}$
and $\bra{0} \Phi_{3,1} \simeq \beta \; \leftidx{_L}{\bra{2}}{}$, where $\beta$ is
some unknown constant.

The trousers trick is used to get rid of $\alpha$ and $\beta$. The
trousers states are again tensor products of two ground states
of size $L/2$ (see Fig. $\ref{fig:latticepants}$). They are all normalized such that the
component of the basis state with empty sites only ($\ket{\;\begin{tikzpicture}
\draw[thick] (0,0) circle (0.01);
\draw[thick] (0.2,0) circle (0.01);
\draw[thick] (0.4,0) circle (0.01);
\draw[thick] (0.6,0) circle (0.01);
\draw[white] (0.2,-0.2) circle (0.001);
\end{tikzpicture}\;}$ or $\bra{\;\begin{tikzpicture}
\draw[thick] (0,0) circle (0.01);
\draw[thick] (0.2,0) circle (0.01);
\draw[thick] (0.4,0) circle (0.01);
\draw[thick] (0.6,0) circle (0.01);
\draw[white] (0.2,0.2) circle (0.001);
\end{tikzpicture}\;}$) is $1$.
Gathering all the pieces of the puzzle, we can define
\begin{equation}
\label{eq:bpolymers}
b(L) \; \equiv \; 4 \; \frac{\left(\leftidx{_L}{\left< \rm{Trousers} |
\tilde{1}\right>}{_L} \right) \left(\leftidx{_L}{\left< \tilde{1} |
\rm{Trousers} \right>}{_L} \right)}{\leftidx{_L}{\left< 1 | \tilde{1} \right>}{_L}} 
\end{equation}
which is independant of all the different choices
of normalization, etc. As in the XXZ chain case ($\ref{eq:bXXZ}$),
we expect that $b(L) \underset{L \rightarrow \infty}{\longrightarrow} b$.
However $b = \left< 0 | L_2 \Phi_{3,1} |0 \right>$ now, and it should
be different from the number $b$ that one finds in the XXZ chain.


\subsection{Numerical results}

We compute $b(L)$ for dilute polymers and the XXZ spin chain at
$q={\rm e}^{i \pi/3}$ (Tab.~\ref{tab:numeric}).  The sizes we can
access are relatively small, therefore our error bars are relatively
large. However they are precise enough to show that $b$ is indeed
different for polymers and the XXZ chain, and compatible with the
predictions of LCFT \cite{GurarieLudwig}: $5/6$ and $-5/8$.

\begin{table}[htbp]
\centering
\begin{tabular}{|c|c||c|c|}
\hline
  \multicolumn{2}{|c||}{\quad Polymers $b(L)$ \quad} &
\multicolumn{2}{c|}{\quad XXZ $b(L)$ \quad} \\
 \hline
Size $L$   & \; $b(L)$ \; &  Size $L$  & \; $b(L)$ \; \\ 
\hline
 2  &  0.68080 &  & \\
 4  &  0.66431 & 4 & -1.36035 \\
 6  &  0.67032 & 8  & -0.87027  \\
 8 &  0.67893 & 12 & -0.75399 \\
 10 &  0.68753 & 16  & -0.70564 \\
 12 &	0.69551 & 	20 & -0.68012 \\
 14 & 0.70273	 &    	&  \\
\hline
 $\infty$ & 0.79$\pm$0.08 &  $\infty$ &  -0.61$\pm$0.02  \\
\hline
\end{tabular}
\caption{Numerical results for the measures of $b(L)$ and extrapolation
  for $L \rightarrow \infty$. The results are compatible with $b=5/6 \simeq 0.833$ for polymers and $b=-5/8 = -0.625$ for the XXZ chain at $q={\rm e}^{i\pi/3}$.}
\label{tab:numeric}
\end{table}

Finally, we stress that fully comparable results are obtained for hamiltonians and for transfer matrices (for the XXZ case for instance, the alternative would involve the 6 vertex model transfer matrix).


\subsection{There is no $b$ in geometrical percolation}

While hamiltonians and  transfer matrices give the same results in the continuum limit, a profound difference can be  observed if one switches representations. That is, instead of the XXZ or 6 vertex model, we could study directly  the  geometrical representation of the percolation problem, either in a loop or in a bond version. In this case we found that 
there  is no Jordan cell at the level of conformal weight two. There are indeed two degenerate fields with energies  (if one works with the geometrical hamiltonian) 
scaling to the conformal weight  $h=2$ (and they coincide with the energies of the XXZ hamiltonian), but the lattice hamiltonian ($L_0$) remains fully diagonalizable.

\paragraph{}
This comes from the structure of the standard representations of
the Temperley-Lieb algebra at $n=1$ (see the discussion below).
In the geometrical representation, both the Hamiltonian $H_L$
and the transfer matrix $T_L$ are built
out of the geometrical generators of the Temperley-Lieb algebra $e_i$'s. For
size $L=4$ these are
$$
\begin{tikzpicture}
	\draw (-0.8,0.26) node{$e_1 \; =$};
	\draw[thick] (0,0.6) arc (-180:0:0.1 and 0.25);
	\draw[thick] (0,0) arc (180:0:0.1 and 0.25);
	\draw[thick] (0.4,0) -- (0.4,0.6);
	\draw[thick] (0.6,0) -- ++(0,0.6);
\end{tikzpicture}
$$
$$
\begin{tikzpicture}
	\draw (-0.8,0.26) node{$e_2 \; =$};
	\draw[thick] (0.2,0.6) arc (-180:0:0.1 and 0.25);
	\draw[thick] (0.2,0) arc (180:0:0.1 and 0.25);
	\draw[thick] (0,0) -- (0,0.6);
	\draw[thick] (0.6,0) -- ++(0,0.6);
\end{tikzpicture}
$$
$$
\begin{tikzpicture}
	\draw (-0.8,0.26) node{$e_3 \; =$};
	\draw[thick] (0.4,0.6) arc (-180:0:0.1 and 0.25);
	\draw[thick] (0.4,0) arc (180:0:0.1 and 0.25);
	\draw[thick] (0,0) -- (0,0.6);
	\draw[thick] (0.2,0) -- ++(0,0.6);
\end{tikzpicture}
$$
For generic $n$, they act on the $6$-dimensional module (we use the same
conventions as in previous sections)
\begin{equation}
\label{eq:basisTL}
\left\{
\ket{\;\begin{tikzpicture}
	\draw[thick] (0,0) arc (-180:0:0.1 and 0.25);
	\draw[thick] (0.4,0) arc (-180:0:0.1 and 0.25);
\end{tikzpicture}\;},
\ket{\;\begin{tikzpicture}
	\draw[thick] (0,0) arc (-180:0:0.3 and 0.25);
	\draw[thick] (0.2,0) arc (-180:0:0.1 and 0.15);
\end{tikzpicture}\;},
\ket{\;\begin{tikzpicture}
	\draw[thick] (0.4,0) arc (-180:0:0.1 and 0.25);
	\draw[thick] (0,0) -- ++(0,-0.25);
	\draw[thick] (0.2,0) -- ++(0,-0.25);
\end{tikzpicture}\;},
\ket{\;\begin{tikzpicture}
	\draw[thick] (0.2,0) arc (-180:0:0.1 and 0.25);
	\draw[thick] (0,0) -- ++(0,-0.25);
	\draw[thick] (0.6,0) -- ++(0,-0.25);
\end{tikzpicture}\;},
\ket{\;\begin{tikzpicture}
	\draw[thick] (0,0) arc (-180:0:0.1 and 0.25);
	\draw[thick] (0.4,0) -- ++(0,-0.25);
	\draw[thick] (0.6,0) -- ++(0,-0.25);
\end{tikzpicture}\;},
\ket{\;\begin{tikzpicture}
	\draw[thick] (0,0) -- ++(0,-0.25);
	\draw[thick] (0.2,0) -- ++(0,-0.25);
	\draw[thick] (0.4,0) -- ++(0,-0.25);
	\draw[thick] (0.6,0) -- ++(0,-0.25);
\end{tikzpicture}\;}
\right\} \; .
\end{equation}
In this basis they can be written
\begin{equation*}
e_1 = \left(  \begin{array}{cccccc}
  n & 1 & 1 & 0 & 0 & 0 \\  
  0 & 0 & 0 & 0 & 0 & 0 \\  
  0 & 0 & 0 & 0 & 0 & 0 \\  
  0 & 0 & 0 & 0 & 0 & 0 \\  
  0 & 0 & 0 & 1 & n & 1 \\  
  0 & 0 & 0 & 0 & 0 & 0 
\end{array} \right)  \quad
e_2 = \left(  \begin{array}{cccccc}
  0 & 0 & 0 & 0 & 0 & 0 \\  
  1 & n & 0 & 0 & 0 & 0 \\  
  0 & 0 & 0 & 0 & 0 & 0 \\  
  0 & 0 & 1 & n & 1 & 1 \\  
  0 & 0 & 0 & 0 & 0 & 0 \\  
  0 & 0 & 0 & 0 & 0 & 0 
\end{array} \right)  \quad
e_3 = \left(  \begin{array}{cccccc}
  n & 1 & 0 & 0 & 1 & 0 \\  
  0 & 0 & 0 & 0 & 0 & 0 \\  
  0 & 0 & n & 1 & 0 & 1 \\  
  0 & 0 & 0 & 0 & 0 & 0 \\  
  0 & 0 & 0 & 0 & 0 & 0 \\  
  0 & 0 & 0 & 0 & 0 & 0 
\end{array} \right)
\end{equation*}
For generic $n \neq 0$ and $n \neq 1$,
the above representation (for $L=4$)
of the Temperley-Lieb algebra is reducible.
Indeed, let us introduce
$$
P = \left( \begin{array}{cccccc}
  1 & 0 & -\frac{1}{n} & 0 & -\frac{1}{n} &	 \frac{n}{(n+1)(n^2-2)} \\  
  0 & 1 & 0 & -\frac{1}{n} & 0 &		 -\frac{1}{(n+1)(n^2-2)} \\  
  0 & 0 & 1 & 0 & 0 &				 -\frac{n-1}{n^2-2} \\  
  0 & 0 & 0 & 1 & 0 &				 -\frac{n-2}{n^2-2} \\  
  0 & 0 & 0 & 0 & 1 &				 -\frac{n-1}{n^2-2}  \\  
  0 & 0 & 0 & 0 & 0 &				 1 
\end{array} \right)
$$
which exists and is invertible as soon as $n \neq 2 \cos \frac{\pi}{k}$,
with $k$ an integer. In general (for larger $L$),
one can construct $P$ using Jones-Wenzl projectors, when they exist. Then
$$
P^{-1} e_1 P = \left(  \begin{array}{cccccc}
  n & 1 &  &  &  &  \\  
  0 & 0 &  &  &  &  \\  
   &  & 0 & 0 & 0 &  \\  
   &  & 0 & 0 & 0 &  \\  
   &  & 0 & 1 & n &  \\  
   &  &  &  &  & 0 
\end{array} \right)  \quad
P^{-1} e_2 P = \left(  \begin{array}{cccccc}
  0 & 0 &  &  &  &  \\  
  1 & n &  &  &  &  \\  
   &  & 0 & 0 & 0 &  \\  
   &  & 1 & n & 1 &  \\  
   &  & 0 & 0 & 0 &  \\  
   &  &  &  &  & 0 
\end{array} \right)$$ $$
P^{-1} e_3 P = \left(  \begin{array}{cccccc}
  n & 1 &  &  &  &  \\  
  0 & 0 &  &  &  &  \\  
   &  & n & 1 & 0 &  \\  
   &  & 0 & 0 & 0 &  \\  
   &  & 0 & 0 & 0 &  \\  
   &  &  &  &  & 0 
\end{array} \right)
$$
and one recognizes the three standard
modules (of dimensions $2$, $3$ and $1$) of the Temperley-Lieb algebra
for $L=4$. Of course, one expects that this generic construction fails
whenever $n$ is a Beraha number $n= 2\cos\frac{\pi}{k}$ with $k$ an integer.
The big surprise here is that the basis change encoded in $P$ is \textit{not}
singular when $n=1$. On the contrary, one gets \textit{two} states that
are annihilated by $e_1$, $e_2$ and $e_3$:
$$
\ket{\;\begin{tikzpicture}
	\draw[thick] (0,0) arc (-180:0:0.1 and 0.25);
	\draw[thick] (0.4,0) arc (-180:0:0.1 and 0.25);
\end{tikzpicture}\;} \; - \;  
\ket{\;\begin{tikzpicture}
	\draw[thick] (0,0) arc (-180:0:0.3 and 0.25);
	\draw[thick] (0.2,0) arc (-180:0:0.1 and 0.15);
\end{tikzpicture}\;} \qquad \rm{and} \qquad 
\ket{\;\begin{tikzpicture}
	\draw[thick] (0,0) -- ++(0,-0.25);
	\draw[thick] (0.2,0) -- ++(0,-0.25);
	\draw[thick] (0.4,0) -- ++(0,-0.25);
	\draw[thick] (0.6,0) -- ++(0,-0.25);
\end{tikzpicture} \;}  \; - \;
\ket{\;\begin{tikzpicture}
	\draw[thick] (0.2,0) arc (-180:0:0.1 and 0.25);
	\draw[thick] (0,0) -- ++(0,-0.25);
	\draw[thick] (0.6,0) -- ++(0,-0.25);
\end{tikzpicture}\;}
$$
and this remains true for larger $L$. Then every operator
built out of the geometrical $e_i$'s (for example $H_L$ and $T_L$)
will have the same structure in this representation: they have
an eigenvalue which is twice degenerate, however it is still
diagonalizable. This is the reason
why there cannot be a Jordan cell at level two in geometrical
percolation. Therefore we had to work in the spin 1/2
representation of the Temperley-Lieb algebra (namely the
XXZ spin chain above), which in the case $n=1$ (\textit{ie} $\Delta=1/2$)
is \textit{not} equivalent to the geometrical representation.

\paragraph{Deformed version of geometrical percolation:} although
the geometrical representation above does not give rise to Jordan cells,
we can deform it slightly to get a geometrical version that is
equivalent to the XXZ chain. This works as follows. For $L=4$, the  
action of the TL generators (recall that $n=1$) in the basis ($\ref{eq:basisTL}$) is now
\begin{equation*}
e_1 = \left(  \begin{array}{cccccc}
  1 & 1 & 1 & 0 & 0 & 0 \\  
  0 & 0 & 0 & 0 & 0 & 0 \\  
  0 & 0 & 0 & 0 & 0 & 0 \\  
  0 & 0 & 0 & 0 & 0 & 0 \\  
  0 & 0 & 0 & 1 & 1 & 1 \\  
  0 & 0 & 0 & 0 & 0 & 0 
\end{array} \right)  \quad
e_2 = \left(  \begin{array}{cccccc}
  0 & 0 & 0 & 0 & 0 & 0 \\  
  1 & 1 & 0 & 0 & 0 & 0 \\  
  0 & 0 & 0 & 0 & 0 & 0 \\  
  0 & 0 & 1 & 1 & 1 & y \\  
  0 & 0 & 0 & 0 & 0 & 0 \\  
  0 & 0 & 0 & 0 & 0 & 0 
\end{array} \right)  \quad
e_3 = \left(  \begin{array}{cccccc}
  1 & 1 & 0 & 0 & 1 & 0 \\  
  0 & 0 & 0 & 0 & 0 & 0 \\  
  0 & 0 & 1 & 1 & 0 & 1 \\  
  0 & 0 & 0 & 0 & 0 & 0 \\  
  0 & 0 & 0 & 0 & 0 & 0 \\  
  0 & 0 & 0 & 0 & 0 & 0 
\end{array} \right)
\end{equation*}
where $y$ is now an arbitrary parameter. It is easy
to check that this defines a representation of the
TL algebra for any $y$, the usual geometrical
representation corresponding to $y=1$. $y$ is a weight
that is given when one contracts the second line (coming from the
infinite past) with the third one. If the first one is
contracted with the second one though (or the third with
the fourth), the weight is $1$. So $y$ is somehow a way of keeping
track of the parity of the lines one has contracted.

We claim that for any $y\neq 1$ this is equivalent
to the XXZ representation. One finds that, if
$$
P_y = \left( \begin{array}{cccccc}
  1 & 0 & 1 & -1 & 0 & -1 \\  
  -1 & 1 & 0 & 0 & -1 & 0 \\  
  0 & 0 & -1 & 1 & 0 & 0 \\  
  0 & 0 & \frac{y-2}{y-1} & 0 & 1 & 0 \\  
  0 & 0 & -1 & 0 & 0 & 1  \\  
  0 & 0 & \frac{1}{y-1} & 0 & 0 & 0 
\end{array} \right)
$$
then
$$
P^{-1}_y e_1 P_y = \left(  \begin{array}{cccccc}
  0 & 1 &  &  &  &  \\  
   & 1 &  &  &  &  \\  
   &  & 0 &  &  &  \\  
   &  &  & 0 & 0 & 0 \\  
   &  &  & 0 & 0 & 0 \\  
   &  &  & 0 & 1 & 1 
\end{array} \right)  \quad
P^{-1}_y e_2 P_y = \left(  \begin{array}{cccccc}
  0 &  &  &  &  &  \\  
   & 1 & 1 &  &  &  \\  
   &  & 0 &  &  &  \\  
   &  &  & 0 & 0 & 0 \\  
   &  &  & 1 & 1 & 1 \\  
   &  &  & 0 & 0 & 0 
\end{array} \right)$$ $$
P^{-1}_y e_3 P_y = \left(  \begin{array}{cccccc}
  0 & 1 &  &  &  &  \\  
   & 1 &  &  &  &  \\  
   &  & 0 &  &  &  \\  
   &  &  & 1 & 1 & 0 \\  
   &  &  & 0 & 0 & 0 \\  
   &  &  & 0 & 0 & 0 
\end{array} \right) .
$$
This clearly shows that for any $y \neq 1$ the deformed
representation has the structure
$$
\begin{tikzpicture}
	\draw (0,1) node{$(1)$};
	\draw[->] (0,0.8) -- (0,0.3);
	\draw (0,0) node{$(1)$};
	\draw[->] (0,-0.2) -- (0,-0.7);
	\draw (0,-1) node{$(1)$};
	\draw (1.6,0) node{$\oplus \qquad (3)$};
\end{tikzpicture}
$$
where the integers are the dimensions of the irreducible representations, and the arrows indicate action of the algebra.
This is exactly the structure expected for the XXZ chain on four sites
at $q=e^{i \pi/3}$ (see the conclusion below).
When $y = 1$, the arrow between the top $(1)$ and the middle $(1)$ is not
there, so we have instead
$$
\begin{tikzpicture}
	\draw (-1.6,0) node{$(1) \qquad \oplus$};
	\draw (0,0.5) node{$(1)$};
	\draw[->] (0,0.3) -- (0,-0.2);
	\draw (0,-0.5) node{$(1)$};
	\draw (1.6,0) node{$\oplus \qquad (3) \; .$};
\end{tikzpicture}
$$
In this sense, the fact that geometrical percolation (as defined
above, i.e., with $y=1$) is actually diagonalizable is an accident.

\paragraph{}
Of course, one can use this deformed representation to compute a
new parameter $b$ as in section $4.1$. The Hamiltonian of the
XXZ spin chain ($\ref{XXZH}$) can be expressed in terms of
the TL generators as $\displaystyle H_L = \frac{L-1}{2} - 2 \sum_{i=1}^{L-1} e_i$
(see also \cite{PasquierSaleur}). Then, proceeding exactly as
in section $4.1$, one can define and compute a lattice $b(L)$
(this involves choosing the scalar product that is compatible with
the action of the algebra on the basis ($\ref{eq:basisTL}$), so this
scalar product involves the parameter $y$ when one contracts the
second and third lines against a half-loop).
We find that this number $b(L)$ does not depend on the choice of $y$
(as soon as $y \neq 1$) and that it is exactly the same
as for the XXZ chain.

\paragraph{}
Meanwhile one could ask what would happen if one studied polymers
in a spin chain or vertex model representation. To answer this,
it is time to move to a slightly more algebraic description.


\section{Conclusion}

Underlying both the XXZ spin chain, the 6 vertex model,  and the geometrical percolation is the Temperley-Lieb algebra, generated by $e_i$ with $i=1,\ldots,L-1$, subject to the famous relations 
\begin{eqnarray}
e_i^2 &=& n \; e_i\nonumber\\
e_ie_{i\pm 1}e_i &=& e_i\nonumber\\
\left[e_i,e_j\right] &=& 0 \quad \mbox{for } |i-j|\geq 2
\end{eqnarray}
with $n=q+q^{-1}$. The Jordan cell structure of the hamiltonian or the transfer matrix is determined by the particular representation of this algebra one is working with: the diagram (geometrical) representation and the vertex representation do not have to behave in the same way.

To state this in more details, we recall that generic irreducible (standard) modules of the  TL algebra are labelled by a number $j$ which is integer or half integer, and have dimensions
\begin{equation}
d_j=\left(\begin{array}{c}
L\\
L/2+j\end{array}\right)-\left(\begin{array}{c}
L\\
L/2+j+1\end{array}\right)
\end{equation}
with the restriction that $L/2+j$ is integer. In the diagram representation, these modules correspond to those where $2j$ strings (or ``through lines'') propagate. In the vertex representation, they correspond to a fixed value $j$ of the $U_q(sl_2)$ spin.

For $q$ a root of unity, and  $j$ such that $[2j+1]_q=0$
these standard modules remain irreducible.%
\footnote{We define the q-analogs as $[x]_q=\frac{q^x-q^{-x}}{q-q^{-1}}$.}
For other values of $j$, these modules contain a proper submodule, but are indecomposable. 
Their structure is independent of $j$: the largest proper submodule is irreducible, and the quotient by this submodule is also. One can represent the structure of such a module by a diagram like
\begin{equation}
\begin{array}{c}
\circ\\
\downarrow\\
\circ\label{stanmod}
\end{array}. 
\end{equation}
Here the top circle represents the states in the simple quotient
module or ``top'' (or ``head''), and the bottom circle the simple
submodule or ``foot''. The arrow represents the action of the
algebra; there is some element of the algebra that maps the top to
the foot, but not vice versa, as well as elements that map the top
into itself and the foot into itself. 

Equivalently, the diagram in (\ref{stanmod}) indicates that, in a basis ordered as (bottom,top),  the Temperley Lieb generators take an upper triangular form.

Of course, since we are dealing with a non semi-simple situation, the particular indecomposables that appear in a physical realization---that is, a representation---can have a rather complicated structure. In the XXZ case, this structure involves further glueing of two standard modules to form a ``diamond'' 
\begin{equation}%
  \begin{array}{ccccc}
       &&\hskip-.2cm\circ&&\\
       &\hskip-.4cm\swarrow\hskip-.2cm&&\hskip-.2cm\searrow&\\
    \circ\hskip-.3cm&&&&\hskip-.3cm \circ\\
       &\hskip-.2cm\searrow\hskip-.2cm&&\hskip-.2cm\swarrow&\\
       &&\hskip-.2cm\circ&&
       \end{array}.\label{tiltmod}
\end{equation}%

Here, 
each circle is a nonzero simple subquotient module, and arrows
show the action of the algebra other than within the simple
subquotients, with the convention that composites of arrows should
also be understood as present implicitly. 

\begin{figure}
 \begin{center}
	\includegraphics[height=60mm]{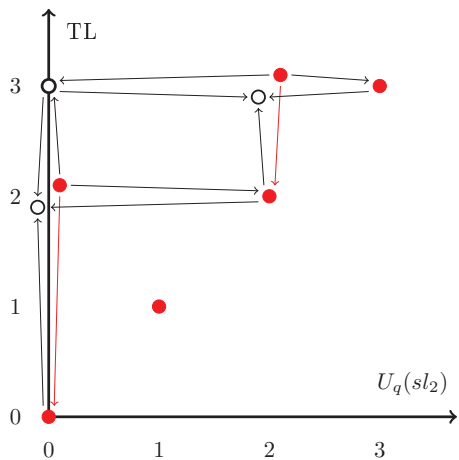}
 \end{center}
  \protect\caption{The structure of the spin-1/2 chain for $q=e^{i\pi/3}$ and $L=6$
  sites, as a representation of $U_q($sl$_2)\otimes$TL$_{L}(q)$.}
  \label{structure}
 \end{figure}

To illustrate this, we borrow from \cite{ReadSaleur} the structure of the XXZ spin chain for $L=6$, as given  in Fig.~\ref{structure}. The horizontal axis contains information about the $U_q(sl_2)$. The vertical axis  encodes information about the TL algebra. Consider what happens above 
the point $0$ of the horizontal axis. The module with top $j=2$ and bottom $j=0$, and the one with top $j=3$ and bottom $j=2$ are the standards, which have gotten glued by further action of the algebra: this is exactly the structure represented in figure \ref{tiltmod} after a $90^o$ rotation.  The two nodes at $j=2$ will go over, in the continuum limit, to two irreducibles of the Virasoro algebra with character $\chi_{1,5}$ (whose state with lowest energy corresponds to $h=2$). The hamiltonian mixes them into rank-two Jordan cells.

The case $L=4$ is similar, although the chain is too short to see the generic structure at $j=0$ - one gets only three simple representations instead of the four ones forming a diamond, as illustrated at the end of the previous section.

In the continuum limit,  it is expected that the diamond goes over a to diamond representation of the Virasoro algebra such as those in \cite{MathieuRidout,KytolaRidout}
\begin{equation}
  \begin{array}{ccccc}
      &&\hskip-.7cmR_{2}&&\\
      &\hskip-.2cm\swarrow&\searrow&\\
      R_{0}&&&\hskip-.3cmR_{3}\\
      &\hskip-.2cm\searrow&\swarrow&\\
      &&\hskip-.7cmR_{2}&&
      \end{array}~~~
      \end{equation}
Here, $R_j$ corresponds to an irreducible of Virasoro with character $\chi_{1,1+2j}$. 
In the $U_q(sl_2)$ spin chain  (the situation would be similar in the SUSY case) there are more multiplicities associated with the extra symmetry. What this means is discussed in \cite{ReadSaleur}. 

Now the point is that the diamond modules do not necessarily have to appear. In fact, in the diagram representation, they do not, at least for values up to $L=8$ \footnote{A general proof most likely exist; this will be addressed elsewhere.}.  In this case, there is no extra multiplicity due to the spin degrees of freedom, and what one gets is a collection of indecomposable  standards with no action of Temperley-Lieb between them. This is represented schematically by the red dots and arrows in Fig.~\ref{structure}. 

In the dilute polymer case, a similar discussion is possible, involving the dilute Temperley Lieb algebra and related geometrical representation, vertex model and spin chain. the decomposition of the Hilbert space for the latter  is given in \cite{ReadSaleur} and is qualitatively similar to Fig.~\ref{structure}. What happens now is that, in the geometrical transfer matrix, one already sees the diamonds, in contrast with the percolation case.

Our findings might have important consequences for the theory in \cite{GurarieLudwig}, although one should be careful in jumping to conclusions. After all, $b$ was introduced and its  value conjectured within a discussion of the chiral sector of bulk properties, and this may not extend straightforwardly to the structure of Virasoro representations observed in the boundary case. 

It is meanwhile a bit surprising that we do not found a Jordan cell for geometrical percolation. We are not sure how this affects the results in \cite{PRZ}. We however note that polymers and percolation are a bit different when considered from the $n\to 0$ resp. $Q\to 1$ limit point of view,   as mentioned briefly in \cite{Cardylog}. 
We will discuss this in more details elsewhere.


\smallskip

\paragraph{Acknowledgments.}

We thank I. Affleck, A.M. Gainutdinov, V. Gurarie,  A.W.W. Ludwig, V. Pasquier, J. Rasmussen and J.M. St\'ephan for discussions.
This work was supported by the Agence Nationale de la Recherche
(grant ANR-06-BLAN-0124-03).




\end{document}